\documentclass[aps,twocolumn,showpacs,preprintnumbers,amsmath,amssymb
,nofootinbib,superscriptaddress,showkeys,prd]{revtex4-1}

\def\XXint#1#2#3{{\setbox0=\hbox{$#1{#2#3}{\int}$}
     \vcenter{\hbox{$#2#3$}}\kern-.5\wd0}}

\usepackage{epsfig}
\usepackage{amsmath}
\usepackage{graphicx}
\usepackage{mathptmx}      
\usepackage{latexsym}
\usepackage{bm}
\usepackage{comment}

\usepackage{latexsym}
\usepackage{amsmath}
\usepackage{amssymb}
\usepackage{amsfonts}
\usepackage{hyperref}
\usepackage{color}

\usepackage{supertabular} 
\usepackage{placeins}
\usepackage{epsfig}
\usepackage{graphicx}

\usepackage{epstopdf}

\usepackage[utf8x]{inputenc}
\usepackage{amsmath}
\usepackage{graphics}
\usepackage{placeins}
\usepackage{epsfig}
\usepackage{hyperref}
\usepackage{color}
\usepackage{diagbox}

\usepackage{soul}

\begin{document} 
\title{On the precise measurement of the $X(3872)$ mass
  and its counting rate
}~\thanks{This work is
  partly supported by the Spanish Ministerio de Economía y
  Competitividad and European ERDF funds (FPA2016-77177-C2-2-P,
  FIS2017-85053-C2-1-P), Junta de Andalucía (FQM-225) and by the EU STRONG-2020 project under the program H2020-INFRAIA-2018-1, grant agreement no.824093.}

\author{Pablo G. Ortega}
\email{pgortega@usal.es}
\affiliation{Departamento de Física Fundamental and \\ Instituto Universitario de F\'isica 
Fundamental y Matem\'aticas (IUFFyM), Universidad de Salamanca, E-37008 
Salamanca, Spain}

\author{Enrique Ruiz Arriola}
\email{earriola@ugr.es}
 \affiliation{Departamento de
  F\'{\i}sica At\'omica, Molecular y Nuclear \\ and Instituto Carlos I
  de F{\'\i}sica Te\'orica y Computacional \\ Universidad de Granada,
  E-18071 Granada, Spain.}

\date{\today}

\begin{abstract}
  The lineshapes of specific production experiments of the exotic
  state such as $X(3872)$ with $J^{PC}=1^{++}$ quantum numbers
  involving triangle singularities have been found to become highly
  sensitive to the binding energy of weakly bound states, thus
  offering in principle the opportunity of benchmark
  determinations. We critically analyze recent proposals to extract
  accurately and precisely the $X(3872)$ mass, which overlook an
  important physical effect by regarding their corresponding
  production lineshapes as a sharp mass distribution and, thus,
  neglecting the influence of initial nearby continuum states in the
  $1^{++}$ channel. The inclusion of these states implies an effective
 cancellation mechanism which operates at the current and finite
  experimental resolution of the detectors so that one cannot
  distinguish between the $1^{++}$ bound-state and nearby $D \bar D^*$
  continuum states with the same quantum numbers. In particular, we
  show that the lineshape for resolutions above 1 MeV becomes rather
  insensitive to the binding energy unless high statistics is
  considered. The very existence of the observed bumps is a mere
    consequence of short distance correlated $\bar D D^*$ pairs, bound
    or unbound. The cancellation also provides a natural explanation
  for a recent study reporting missing but unknown decay channels in
  an absolute branching ratio global analysis of the $X(3872)$.
\end{abstract}



\maketitle

\section{Introduction}

The quest for the hadronic spectrum has been a major goal in particle
physics over the last 70 years, which has been marked by predicting and
reporting the observed states and their properties in the PDG (see
e.g. ~\cite{PDG} for the latest edition upcoming).  Before 2003, this
task has mostly been phenomenologically supported by a
non-relativistic quark model pattern and its given symmetry multiplets
suggested by the underlying $q\bar q$ and $qqq$ composition for mesons
and baryons, respectively.  This non-rigorous but effective link has
been a quite useful and extremely relevant guidance, particularly
because, currently, it is theoretically unknown how many states
should occur below a given maximal energy or if the full set of
recorded states are incomplete or
redundant~\cite{RuizArriola:2016qpb}. In fact, as it is most often the
case for hadronic resonances, we do not detect directly the reported
particle through its track but only in terms of its decaying products
so that the corresponding invariant mass distribution is observed
instead and the relevant signal is singled out from the reaction
background within a {\it given} energy resolution.

\begin{figure}[th]
\includegraphics[width=.45\textwidth]{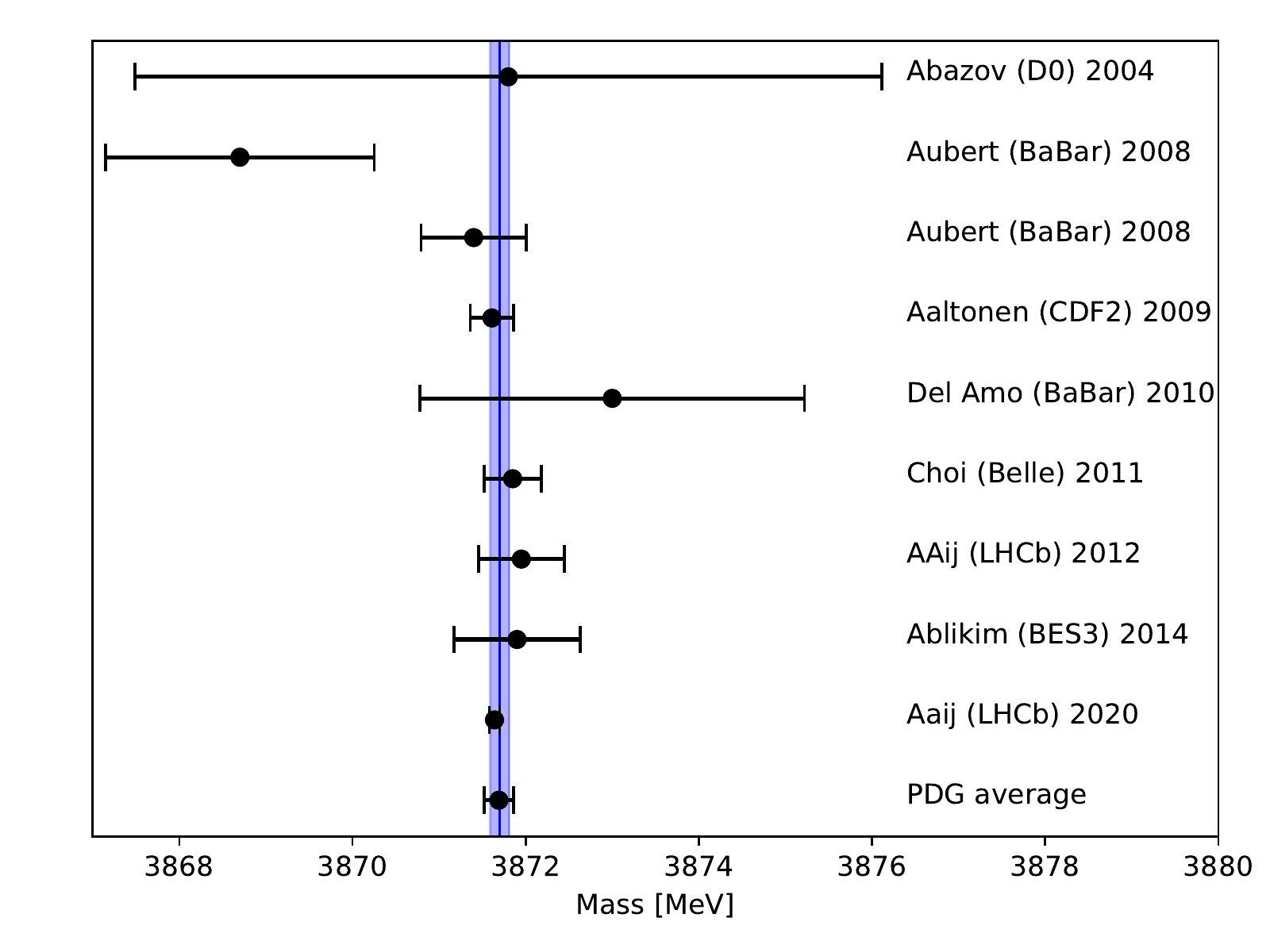}
\caption{\label{fig:X-Prob} Different $X(3872)$ masses
  determinations~\cite{Choi:2003ue,Aubert:2004zr,Choi:2011fc,Aaij:2013zoa}
  with standard $68\%$ confidence limits. The band corresponds to the
  current $\bar D D^*$ threshold value with uncertainties.}
\end{figure}

Since 2003, the situation has become more involved above charm
production threshold after the discovery of the
$X(3872)$~\cite{Choi:2003ue,Aubert:2004zr,Choi:2011fc,Aaij:2013zoa}
and the wealth of new $X,Y,Z$ states whose properties suggest more
complicated structures than those originally envisaged from the
quark-model~\cite{Godfrey:2008nc,Guo:2017jvc,Brambilla:2019esw}. In
this study, we analyze the renowned $X(3872)$ state and the influence
of the mass distribution in the $1^{++}$ channel on the determination
of its mass. The $X(3872)$ is allegedly a $\bar D D^*$ weakly bound
state, whose binding energy has become smaller since its
discovery. The most recent value for its binding energy, measured by
LHCb, is 0.07(12) MeV~\cite{Aaij:2020xjx} for a $1\sigma$
  confidence level. This actually corresponds to a $30\%$ probability
  of {\it not} being a bound state. We illustrate the situation in
  Fig.~\ref{fig:X-Prob} within a conventional Gaussian distribution
  profile interpretation. So, at present, it is unclear whether its
  mass is slightly above or below the $\bar D D^*$ threshold. 
However, one might wonder what would happen if the X(3872) is not
 a bound state. Recently several proposals invoke the strong
sensitivity of lineshapes for production processes involving triangle
singularities to benchmark the mass determination~\cite{Guo:2019qcn,Braaten:2019gfj}.

In this paper, we promote the idea that the precise value of the mass
is actually not crucial, since the contribution of nearby states with
the {\it same} quantum numbers is unavoidable with the current
experimental energy resolution detecting its decaying products, and a
cancellation mechanism put forward initially by Dashen and
Kane~\cite{Dashen:1974ns} is at work in this particular case.  We have
found in previous works that this has implications to count $X(3872)$
degrees of freedom at finite temperatures of relevance in relativistic
heavy ions collisions~\cite{Ortega:2017hpw,Ortega:2017shf} and
ultrahigh energies pp prompt $X(3872)$ production at finite $p_T$ and
mid-rapidity~\cite{Ortega:2019fme}.  We will also show how the number
of reconstructed states representing the bound $X(3872)$ is {\it
  smaller} than the truly produced ones due to a cancellation
mechanism which will be explained below and which provides a natural
understanding of the missing decay channels. A brief account and
overview of the present study has already been advanced in conference
proceedings~\cite{RuizArriola:2020ijw}.

The paper is organized as follows: In section \ref{sec:dos} we review
the hadronic density of states and its theoretical and experimental
limitations as it will be a key element of our analysis. In section
\ref{sec:XYZ} we review the XYZ states to provide a broader
perspective around the very special $X(3872)$ exotic state. In section
\ref{sec:X3872} we approach the determination of the density of states
in the $1^{++}$ channel. Our main numerical results are discussed in
section \ref{sec:num}.  In section \ref{sec:corr} we ponder on the
relevance of $D \bar D^*$ correlations, rather than binding, as the
key behind the observed signals.  Finally, our conclusions are
presented in Section~\ref{sec:concl}.

\section{Hadronic density of states}
\label{sec:dos}

\subsection{General properties}

For completeness, in this section we review some basic aspects of the hadronic
density of states following some historical timeline, in a way that
our points can be more easily presented and with the purpose of fixing the
notation. The first quantum-mechanical attempt to determine the
density of states within the quantum virial expansion was pioneered by
Beth and Uhlenbeck in 1937, who computed the second virial coefficient
as a function of temperature in terms of the two-body scattering phase
shifts~\cite{Beth:1937zz}. Only after 30 years, Dashen, Ma and
Bernstein provided, in a seminal work, the link to the full S-matrix
\cite{Dashen:1969ep} which opened up the basis for the Hadron
Resonance Gas (HRG) model for resonances~\cite{Dashen:1974jw}, as well as the notion of effective elementarity~\cite{Dashen:1974yy}. Based on
these developments, Dashen and Kane promoted the natural idea of
counting hadronic states at a typical hadronic scale. In terms of the
corresponding density of states as a function of the invariant CM
energy $\sqrt{s}$~\cite{Dashen:1974ns}, we have
\begin{eqnarray}
  \rho (M) = {\rm Tr} \delta(M-H_{\rm CM}) = \sum_n \delta (M-M_n)
\label{eq:rho(M)}  
\end{eqnarray}
where $H_{\rm CM}$ is the intrinsic Hamiltonian and $M_n$ the
  corresponding eigenvalues. We use here a bound state notation but, in
  practice, the continuum spectrum which will be of concern here implies
  a spectral integral which can be approximated by imposing a
  discretization approximation, such as placing the system on a
  sufficiently large box with finite volume. Unfortunately, while
this is mathematically a well-defined quantity, $\rho(M)$ cannot, in
most cases, be computed or measured directly, but only through its
coupling to external probes generating the production process. This
effectively correspond to multiply by an observable ${\cal O}(M)$
 and superimpose the contributions in a given energy
  window. Another possibility is the coupling to a thermal heat bath
where we take this observable to be a universal Boltzmann factor
$e^{-M/T}$.

\subsection{The two-body case}

The level density can be splitted into separate contributions
according to the corresponding good quantum numbers. In the particular
$2 \to 2 $ process (for a recent discussion of N-body and coupled
channel aspects see e.g. Refs.~\cite{Lo:2017sde,Lo:2020phg} and
references therein) one has that the interacting cumulative number in a
given channel in the continuum with threshold $M_{\rm th}$ is given
as~\cite{Fukuda:1956zz,DeWitt:1956be} (for updated presentations see 
e.g.~\cite{Gomez-Rocha:2019xum,Gomez-Rocha:2019rpj})
\begin{eqnarray}
  \Delta N(M) & \equiv & N(M)-N_0(M) \nonumber \\ 
  &=& \sum_n \theta(M-M_n^B)
  + \frac1\pi \sum_{\alpha=1}^{n} [\delta_\alpha (M)-\delta_\alpha (M_{\rm th})] \, .
\label{eq:ncum} 
\end{eqnarray}
Here, we have separated bound states $M_n^B$ explicitly from scattering
states written in terms of the eigenvalues of the S-matrix, i.e. $S =
U {\rm Diag} (\delta_1, \dots, \delta_n ) U^\dagger$, with $U$ a
unitary transformation for n-coupled channels. This definition fulfills
$N(0)=0$. In the single channel case, and in the limit of high masses
$M \to\infty$ one gets $ N(\infty)=n_B + \frac1{\pi}
[\delta(\infty)-\delta(M_{\rm th})]=0 $ due to Levinson's theorem. The
opening of new channels and the impact of confining interactions was
discussed in Ref.~\cite{Dashen:1976cf}. According to Dashen and Kane,
some states may present a fluctuation at the hadronic scale so that
their contribution cancels, so that the state {\it does not} count.

\subsection{Theoretical binning}

From a purely theoretical side, a practical and numerical evaluation
 of the level density rests on the computation of the energy
  levels, $M_n$, as demanded by Eq.~(\ref{eq:rho(M)}) which could, in
  principle, be evaluated with arbitrary precision. In practice, 
  this evaluation requires binning
the spectrum with a given finite invariant mass resolution $\Delta m$,
in which case only an averaged or coarse-grained value such as~\cite{Dashen:1974ns}
\begin{eqnarray}
\bar \rho (M) = \frac1{\Delta m} \int_{M-\Delta m/2}^{M+\Delta m/2}
\rho(m) dm
\label{eq:rho-av}
\end{eqnarray}
is obtained. On the theoretical side, a practical way of implementing
this is by placing the system into a box of volume $V$, as it is the
case in lattice QCD where one roughly has $\Delta m \sim V^{-1/3}$.
This finite mass resolution effectively corresponds to a coarse
graining in mass and should not have any sizable effect on the
result, {\it unless} the true density of states presents large
fluctuations on a smaller mass scale. With this viewpoint in mind,
Dashen and Kane made the distinction between the original $SU(3)$
multiplets and ``accidental'' states, i.e.  those states which {\it do
  not} contribute when $\Delta m$ is sufficiently large (presumably
about the typical symmetry breaking multiplet splitting).

\subsection{Experimental resolution}

On the experimental side, the coarse-graining procedure corresponds to
the finite energy resolution of the detectors, typically
$\sigma=1-3$MeV (see also the discussion below).  The
amount of inherent fluctuation is estimated by assuming that the formation of each
charge carrier in the detector is a Poisson process. This average
corresponds to use a Gaussian detector response function with
$\sigma$-broadening,
\begin{equation}
 R_\sigma(m,M)=\frac{1}{\sqrt{2\pi}\sigma} e^{-\frac{(m-M)^2}{2\sigma^2}}
\end{equation}
so, we
have~\cite{knoll2010radiation}
\begin{eqnarray}\label{eq:smearedrho}
 \bar{\rho}_\sigma(M) =\int_{-\infty}^{\infty} R_\sigma (m,M)\rho(m) dm 
\end{eqnarray}
The binning procedure implied by Eq.~(\ref{eq:rho-av}) may be added
afterwards. Although it is innocuous for $\Delta m \le \sigma$, 
it can have a sizable effect for $\Delta m > \sigma$.

\subsection{The Dashen-Kane cancellation}

The immediate consequence of the particular phase shift behavior
follows from Eq.~\ref{eq:ncum} at the density of states level, defined
as
\begin{eqnarray}
  \rho(M) = \frac{d \Delta N(M)}{dM}=
\sum_n \delta(M-M_n^B)
  + \frac1\pi \sum_{\alpha=1}^{n} \delta'_\alpha (M) \, .
\label{eq:dndm}
\end{eqnarray}

Assuming an experimental resolution $R_\sigma(m,M)$, the corresponding measured quantity for an observable depending on the invariant mass function
$O(M)$ is

\begin{align}
 O_{\rm meas} (M)  &=\int_{-\infty}^{\infty} O (m) R_\sigma (m,M)\rho(m) dm.
\end{align}

Then, for a bin in the range
$(M-\Delta m/2 , m+\Delta m/2)$, it becomes
\begin{eqnarray}
  O_{\rm meas} \equiv \frac{1}{\Delta m}\int_{M-\tfrac{\Delta m}{2}}^{M+\tfrac{\Delta m}{2}} O_{\rm meas}(M') dM' .
\label{eq:Obs}
\end{eqnarray}

In the single channel case, with phase shift $\delta(M)$, 
one has 
\begin{align}
 O_{\rm meas} &={\cal R}(M^B) O(M^B)+\frac{1}{\pi}\int_{-\infty}^{\infty} {\cal R}(m)O(m)\delta'(m) dm\,,
 \label{eq:Obs-can}
\end{align}
with  ${\cal R}(m)=\frac{1}{2\Delta m}\left[{\rm Erf}\left(\frac{m-M^B+\Delta m/2}{\sqrt{2}\sigma}\right)+{\rm Erf}\left(\frac{M^B-m+\Delta m/2}{\sqrt{2}\sigma}\right)\right]$.
Which, for a decreasing phase-shift and for a smooth observable
$O(M)$, points to a cancellation whose precise amount depends on the
corresponding slope above threshold.

\subsection{The deuteron state and the np continuum}

The cancellation between the continuum and discrete parts of the
spectrum was pointed out by Dashen and Kane long
ago~\cite{Dashen:1974ns} (see also
\cite{Arriola:2015gra,Arriola:2014bfa} for an explicit picture and
further discussion within the HRG model framework).  A prominent
example of such a cancellation discussed in these works
corresponds to the deuteron, which is a neutron-proton $1^{++}$ state
weakly bound by $B_d=2.2 {\rm MeV} \ll m_p+m_n \sim 1980 {\rm
  MeV}$. This effect can explicitly be observed in the np virial
coefficient at rather low temperatures~\cite{Horowitz:2005nd} (
  this work however fails to link the effect to the Dashen-Kane
  effect). While this cancellation is not exactly a theorem, it is an
open possibility {\it a fortiori} whose verification depends on
details of low energy scattering. We point out that the cancellation
observed in the equation of state for nuclear matter at low
temperatures where one has a superposition of states weighted by a
Boltzmann factor~\cite{Horowitz:2005nd} corresponds to a suppression
of the occupation number in the $1^{++}$ channel as compared to the
deuteron case, $N_{1^{++}} \le N_d $.

The case of the deuteron described above is particularly interesting
for us here since it is extremely similar to the case of the $X(3872)$,
with the important exception of the detection method of both states,
as will be discussed below. In our previous work~\cite{Ortega:2017hpw}
we have shown how this cancellation can likewise be triggered at
finite temperature $T$ for the $X(3872)$, as it is the case in
relativistic heavy ion collisions, since the partition function
involves the folding of the Boltzmann factor, $ \sim
e^{-\sqrt{p^2+m^2}/T}$ with the density of states,
Eq.~\ref{eq:dndm}. Therefore, given these suggestive similarities, we
have undertaken a comparative study of the deuteron {\it and}
$X(3872)$ production rates in pp scattering at ultra-high energies
($\sim 7$ TeV) in the observed $p_T$ distributions in colliders, which
provides a suitable calibration tool in order to see the effects of
the cancellation due to the finite resolution $\Delta m$ of the
detectors signaling the $X(3872)$ state and deciding on its bound state
character~\cite{Ortega:2019fme}.

\section{The XYZ states}
\label{sec:XYZ}

Nowadays, there is a strong theoretical and experimental evidence on the
existence of loosely bound states near the charm threshold, originally
predicted by Nussinov and Sidhu~\cite{Nussinov:1976fg}, as it seems to be
confirmed now by the wealth of evidence on the existence of the
$X(3782)$, re-named  $\chi_{c1} (3872)$,  
state with binding energy
$B_X=0.01(18)$MeV~\cite{Tanabashi:2018oca}, or  $0.07(12)$ MeV from recent LHCb measurements~\cite{Aaij:2020xjx}, and which has triggered a
revolution by the proliferation of the so-called X,Y,Z states (for
reviews see
e.g.~\cite{Esposito:2016noz,Guo:2017jvc,Brambilla:2019esw}.  In the
absence of electroweak interactions, this state has the smallest known
hadronic binding energy and, for a loosely bound state, many properties
are mainly determined by its binding energy~\cite{Guo:2017jvc} since
most of the time the system is outside the range of the
interaction.

In fact, the molecular interpretation has attracted
considerable attention, but since this state is unstable against
$J/\psi \rho$ and $J/\psi \omega$ decays, the detection of $X(3872)$
relies on its decay channels spectra where the mass resolution never
exceeds $\Delta m \sim 1$-$2
$MeV~\cite{Choi:2003ue,Aubert:2004zr,Choi:2011fc,Aaij:2013zoa} (see
e.g. \cite{Karliner:2017qhf} for a graphical summary on the current
spectral experimental resolutions). Therefore it is in principle
unclear if one could determine the mass of the $X(3872)$ or,
equivalently, its binding energy $\Delta B_X \ll \Delta m$ with such a
precision, since we cannot distinguish sharply the initial state.
While in most studies (see however \cite{Kang:2016jxw}) the bound
state nature is assumed rather than deduced, even if the $X(3872)$ was
slightly unbound the correlations would be indistinguishable in the
short distance behavior of the $D \bar D^{*0}$ wave function.

The discussion on $X(3872)$ lineshapes started in
Ref.~\cite{Braaten:2007dw} as a way to extract information on the
binding. Triangle singularities are ubiquitous in weakly bound
hadronic and nuclear systems~\cite{Karplus:1958zz} and arise when
three particles in a Feynman diagram can simultaneously be on the
mass shell. Their relevance in XYZ states has been pointed
out~\cite{Szczepaniak:2015eza} and their relation to unitarity has
been emphasized~\cite{Oset:2018bjl,Liu:2015taa}. In fact, they have
been put forward recently as a method to sensitively determine the X
mass based on the theoretical line shape.  The fall-off of the
lineshape above the peak, rather than the actual position of the peak
reflects rather well the binding energy~\cite{Guo:2019qcn,Braaten:2019gfj,Sakai:2020ucu}.

F.-K.~Guo has considered the effect of a short distance source (the specific
process has not been specified) which generates a $D^{*0} \bar D^{*0}$
pair in a relative S-wave and which eventually evolves into a
$X(3872)+ \gamma $ final state~\cite{Guo:2019qcn}. This production
mechanism is enhanced by the $D^{*0} \bar D^{*0} \to \gamma D^0 + \bar
D^{*0} \to \gamma + X(3872)$ one loop triangle singularities producing
a narrow peak at about the $D^{*0} \bar D^{*0}$ threshold. E.~Braaten,
L.-P. He and K.~Ingles have proposed a similar triangle singularity
enhancement for the production of X(3872) and a photon using $e^+e^-$
annihilation as the source of a $D^{*0} \bar D^{*0}$ pair in a
relative P-wave, which becomes possible because of its $1^{++}$ quantum
numbers~\cite{Braaten:2019gfj}. Further related analysis on this
regard may be found in Ref.~\cite{Sakai:2020ucu,Molina:2020kyu}.

However, these methods focusing on the $X(3872)$ production lack one
important circumstance operating due to the finite resolution of the
detectors, since they assume a {\it pure} initial mass state (mostly
the bound state mass $M_X$). In reality, any nearby initial states with
the {\it same} $1^{++}$ quantum numbers will produce a signal in the
final state due to the finite resolution in the final state.  We have
reported recently on the neat and accurate cancellation between the
would-be X(3872) bound state and the $D \bar D^*$ continuum in the
initial state which has a sizable impact on the final density of
states and blurs the detected
signal~\cite{Ortega:2017hpw,Ortega:2017shf}. In this work, we will
extend those works to analyze the implications on the allegedly
accurate mass determinations.

The similarities between $d$ and $X(3872)$ already noted in
Refs.~\cite{Tornqvist:1993ng,Close:2003sg,Braaten:2003he} have been
corroborated on a quantitative level in our recent
work~\cite{Ortega:2019fme}, were we have pointed out that they are also
applicable from the point of view of production at
accelerators~\cite{Ortega:2019fme}. However, a crucial and relevant
difference for the present work is that while the deuteron is detected
{\it directly} by analyzing its track and/or stopping power leaving a
well-defined trace, the $X(3872)$ is inferred from its decay
properties, mainly through the $J/\psi \rho $ and $J/\psi \omega $
channels.

\section{Level density in the $X(3872)$ channel}
\label{sec:X3872}

\subsection{Coupled channel scattering}

\begin{figure}[h]
\includegraphics[width=.45\textwidth]{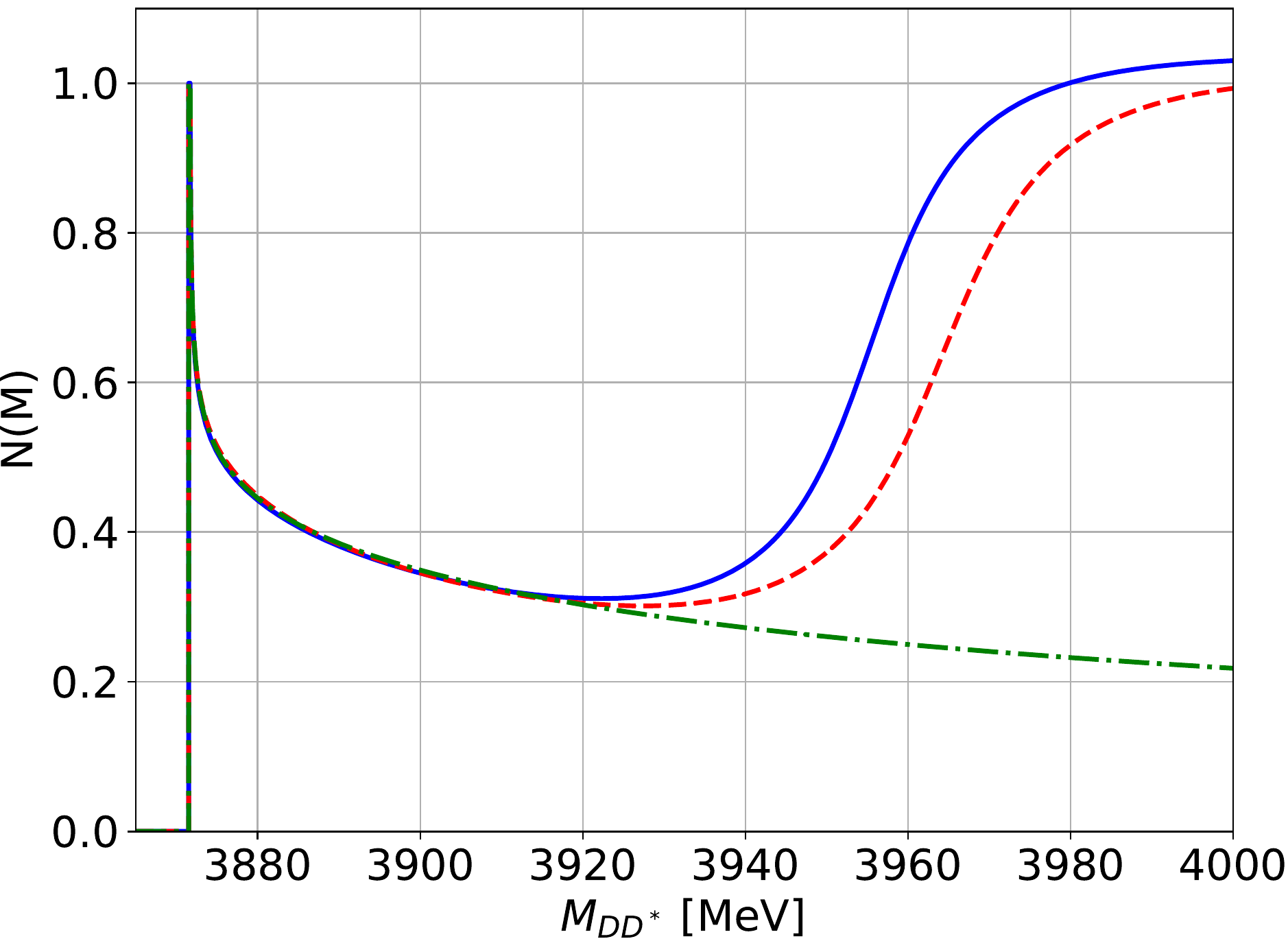}
\caption{\label{fig1} Comparison between the cumulative number of the
  $1^{++}$ sector with $E_b=180$ keV in different models: The
  coupled-channels EFT model of Ref.~\cite{Cincioglu:2016fkm} with
  $d=0.4$ fm$^{1/2}$, $C=-976$ fm$^2$ and $m_{c\bar c}^{(0)}=3947.44$
  MeV (blue); the coupled-channels CQM model of
  Ref.~\cite{Ortega:2009hj} with $m_{c\bar c}^{(0)}=3947.44$ MeV and
  $\gamma_{^3P_0}=0.194$ (dashed red) and the Effective Range Approximation (ERA) model with $r_0=1$ fm and
  $a_s=\frac{1}{\sqrt{2\mu E_b}}=10.58$ fm (dash-dot green).}
\end{figure}

In order to implement the formula given by Eq.~(\ref{eq:ncum}), we make
some digression on the $D \bar D^*$ scattering states in the $1^{++}$,
which actually resembles closely the same channel for the deuteron.
However, while the partial wave analysis of NN scattering data and the
determination of the corresponding phase-shifts is a well-known
subject, mainly due to the abundance of data~\cite{Perez:2013jpa}, we
remind that a similar analysis in the $D \bar D^*$ case is, at present,
in its infancy and thus our first analysis in
Ref.~\cite{Ortega:2017hpw} has been based on a quark-model.  In the
$1^{++}$ channel, the presence of tensor force implies a coupling
between the $^3S_1$ and $^3D_1$ channels, so that the S-matrix is
given by
\begin{eqnarray}
S^{J1} &=&\left(
\begin{array}{cc}
\cos \epsilon_j & -\sin \epsilon_j \\ \sin
\epsilon_j & \cos \epsilon_j 
\end{array}
\right) \left(
\begin{array}{cc}
e^{2 {\rm i}
\delta^{1j}_{j-1}} & 0 \\ 0 & e^{2 {\rm i} \delta_{j+1}^{1j}} 
\end{array}
\right) \nonumber \\
&\times& 
\left( \begin{array}{cc}
\cos \epsilon_j & -\sin \epsilon_j \\ \sin
\epsilon_j & \cos \epsilon_j 
\end{array} 
\right) \, . 
\end{eqnarray}
From here we define the T-matrix
\begin{equation}
S^{JS}= 1 - 2 i k T^{JS} \, ,    
\end{equation}
The S and D eigen phase-shifts have been shown in our previous
work~\cite{Ortega:2017hpw} using the quark cluster model of
Ref.~\cite{Ortega:2009hj,Ortega:2012rs} which includes both a $c \bar c$ and $D \bar
D^*$ channels. The cumulative number is shown in Fig.~\ref{fig1}. The
outstanding feature is the turnover of the function as soon as a
slightly non-vanishing $c \bar c$ content in the $X(3872)$ is
included, unlike the purely molecular picture (see
Ref.~\cite{Ortega:2017hpw} for a more detailed discussion). We also
compute the cumulative number for the coupled-channels EFT model of
Ref.~\cite{Cincioglu:2016fkm} fine-tuning the parameters to agree at
low energies with the quark model. In both
cases the fitting parameters have been binding properties of the
$X(3872)$.  As we see, results present a
rather similar pattern over the entire plotted energy range; the
sharp rise of the cumulative number is followed by a strong decrease
generated by the phase-shift. Moreover, we have checked that the
S-wave phase-shift asymptotically approaches $\pi$ (due to the bound
X(3940)-state of the purely confined channel~\cite{Ortega:2012rs}
which becomes a resonance when coupled to the $D \bar D^*$ continuum)
and hence $N(\infty)=1$ in agreement with the modified Levinson's
theorem of interactions with confining channels~\cite{Dashen:1976cf}.

\subsection{Effective range approximation}

However, as we will see, the S-D waves mixing stemming from the tensor
force has an influence for larger energies than those considered
here~\cite{Ortega:2017hpw}. Therefore, in order to illustrate how the
cancellation comes about, we also considered a simple model which
works fairly accurately for {\it both} the deuteron and the $X(3872)$
by just considering a contact (Gaussian)
interaction~\cite{Gamermann:2009uq} in the $^3S_1$-channel and using
effective range parameters to determine the corresponding
phase-shift in the $d$ and
$X(3872)$~\cite{Arriola:2013era,Ortega:2017hpw} respectively. The
result for $N(M)$ together with the EFT and CQM predictions can be
seen in Fig.~\ref{fig1}. Of course, if the binding energy is not that
small, several effects appear and, in particular, the composite nature
of the $X(3872)$ becomes manifest (see e.g.~\cite{Ortega:2009hj}). All
these similarities suggests the possibility of using the
shape-independent Effective Range Approximation (ERA) to second order
to calculate the phaseshifts near threshold.  In ERA, we have that the
$\delta$ is given as a function of two parameters:
\begin{equation}
 k\, \cot\, \delta = -\frac{1}{a_s}+\frac{1}{2}r_0 k^2
\end{equation}
where $k$ is the CM momentum 
\begin{eqnarray}
k=\sqrt{2\mu (M-M_0)}
\end{eqnarray}
where $\mu = M_D M_{D⁺} /(M_D +M_{D*})$ is the reduced mass and
$M_0=M_D+M_{D*}$ is the threshold mass. The comparison in
Fig.~\ref{fig1} the between ERA and the two coupled-channels models
reassures the validity of the approximation for the range $\sqrt{s}
\lesssim 3920$MeV.
\begin{figure}[ttt]
 \includegraphics[width=.45\textwidth]{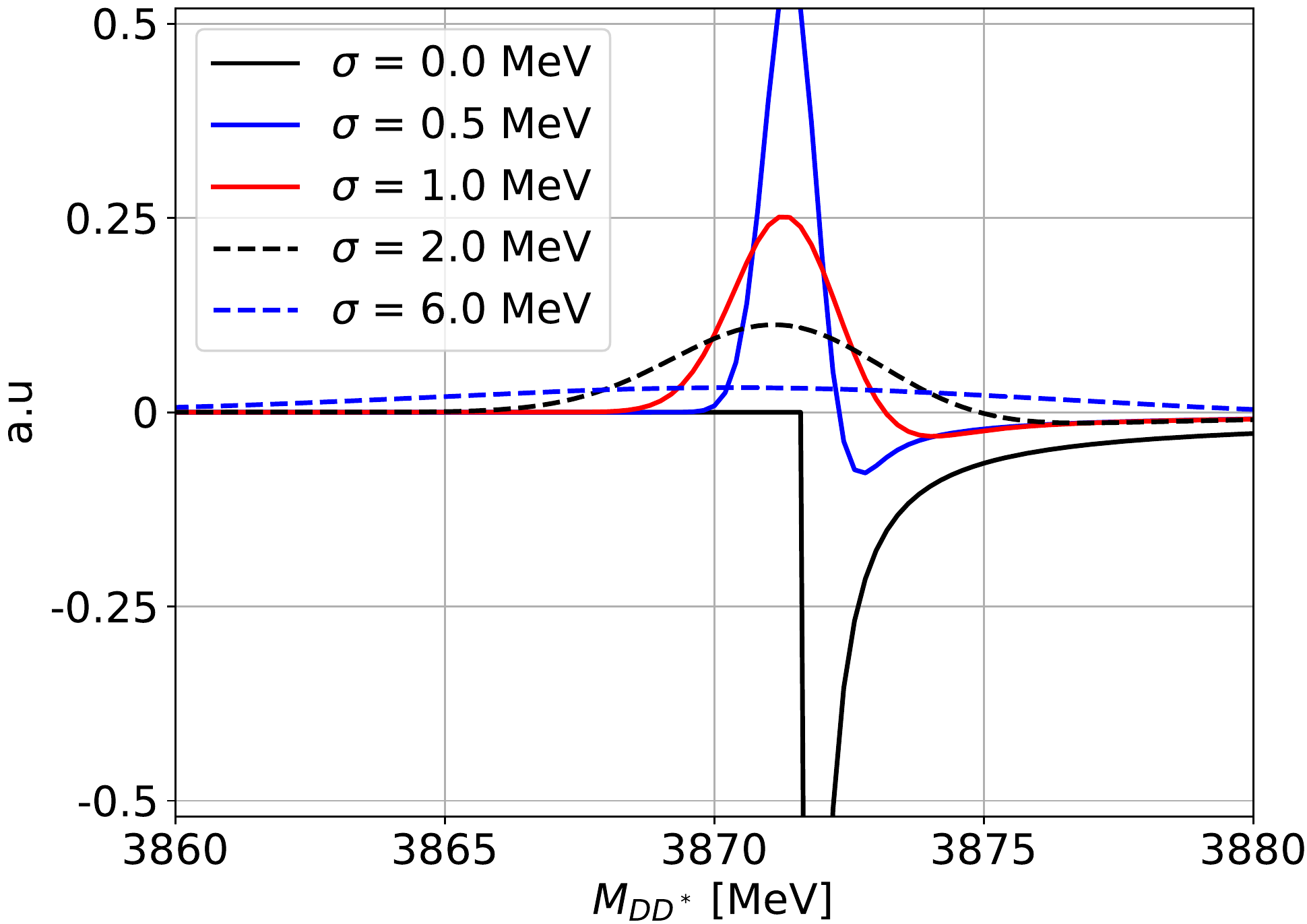}
 \includegraphics[width=.45\textwidth]{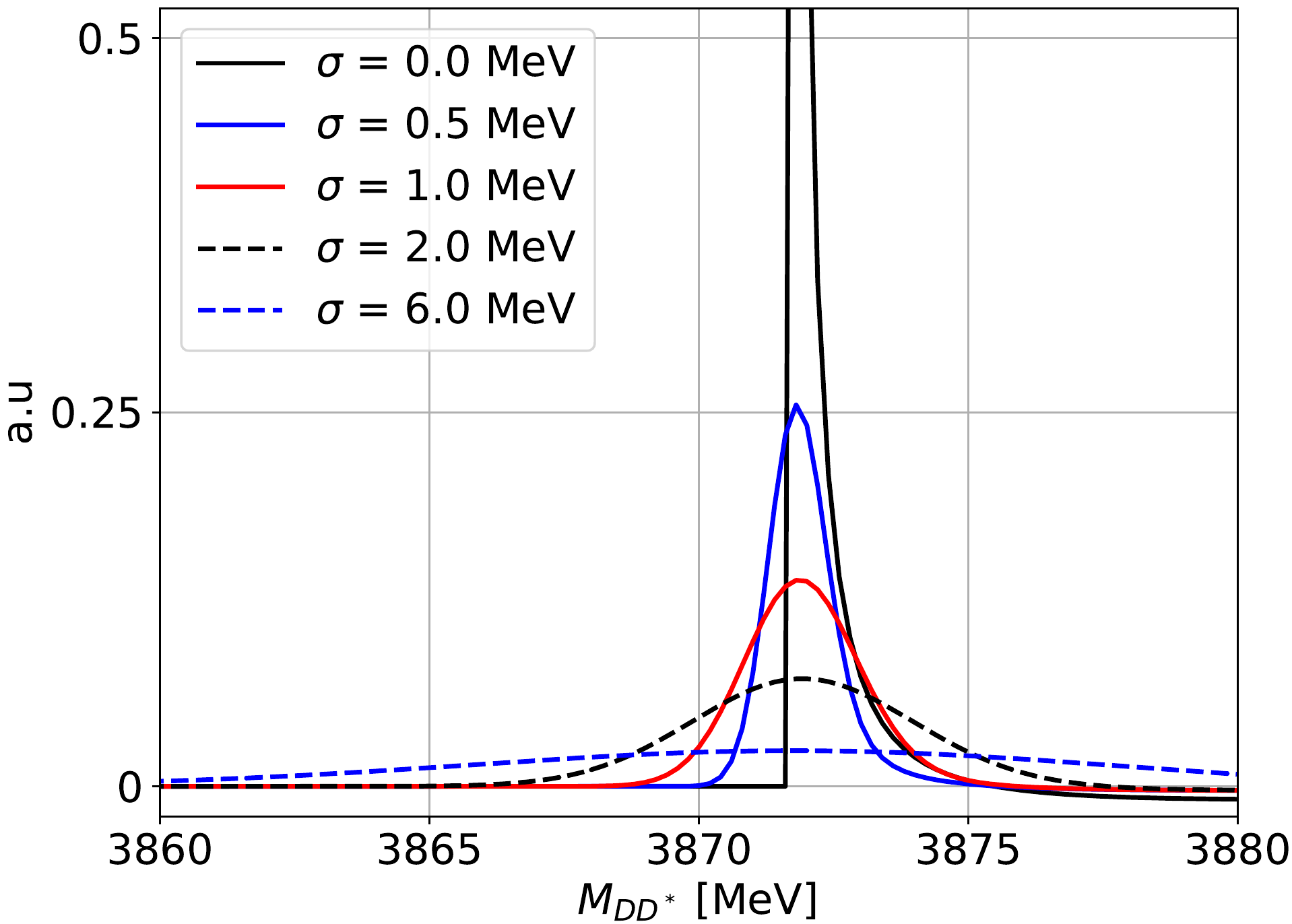}
 \caption{\label{fig2} Upper: Smeared density of states for $E_b=180$ keV for different resolutions. Lower: Same for $E_b=-180$ keV (virtual).}
\end{figure}
The partial wave inverse scattering amplitude is given by
\begin{eqnarray}
f_0(k)^{-1}= k \cot \delta - i k    
\end{eqnarray}
and, in general, bound and virtual states correspond to poles of
$f_0(k)$ at $ k = \pm i \gamma_X$ in the first and second Riemann
sheet in energy $E_b=M_X-M_0$ respectively. It is worth mentioning
that Kang and Oller have comprehensively studied the pole structure and
analyzed the character of the $X(3872)$ in terms of bound and virtual
states within simple analytical parameterizations~\cite{Kang:2016jxw},
although the Dashen-Kane cancellation was not addressed.

\subsection{Finite energy resolution}

The detector response function transforms the monochromatic signal of
mass $M_X$ in a Gaussian distribution $R_\sigma(M_X,m)$ with $\sigma$
resolution~\cite{knoll2010radiation}. It reflects the imperfection of
the detector to measure a single energy due to the Poisson statistics
of the energy deposition. The energy window $\Delta M$ is interpreted
as the energy range where the final channel products are selected as
decay products of the $X(3872)$ (and, thus, reconstructed). Usually
they are taken as $\pm (2-3)\sigma$, to take most of the Gaussian
distribution. The binning energy $\Delta m$ corresponds to the actual
sampling of discriminated data.

The experiments measure such Gaussian distributions, from where the
typical resolution $\sigma$ can be extracted. For example, in
Ref.~\cite{Ablikim:2013dyn}(page 4) the authors claim a resolution of
$\sigma\approx 1$ MeV when measuring the mass of the $\psi(3686)$, and
$J/\psi\pi\pi$ events between $3.86$ and $3.88$ GeV are selected,
thus, employing an energy window of $20$ MeV and binning with $3$
MeV. The situation for this and other
experiments~\cite{Choi:2011fc,Aaij:2011sn} is summarized in Table
\ref{tab:1} (see appendix~\ref{sec:appendix} for details).

\begin{table}
\begin{tabular}{c|c|c|c|c}
\hline\hline
Channel & $\sigma$ &  $\Delta m$ & $\Delta M$ & Reference \\
\hline
$J/\psi\pi^+\pi^-$ & $1.14\pm0.07$ & 3 & $20$ & Ref.~\cite{Ablikim:2013dyn} \\
$J/\psi\pi^+\pi^-$ & $3.33\pm0.08$ & 2 & $6\sigma\approx 20$ & Ref.~\cite{Aaij:2011sn} \\
$J/\psi\pi^+\pi^-$ & $4$ & $2$& $18$ & Ref.~\cite{Choi:2011fc} \\
\hline
\end{tabular}
\caption{\label{tab:1} Detector energy resolutions $\sigma$, binning
  $\Delta m$ and energy window $\Delta M$ in several experiments
  detecting $X(3872)$ decays.}
\end{table}

According to Table \ref{tab:1} the finest value for the resolution
$\sigma$ is around $\sigma=1$ MeV, and it reflects the best possible experimentally
accessible resolution at present. Additionally, the energy window of 
selected events would be
of $\Delta M= 20$ MeV.

\subsection{Smearing of the density of states}

According to the general expression, Eq.~\ref{eq:dndm}, and neglecting
the inessential S-D wave mixing at low energies, the density of states
in the $1^{++}$ channel for the bound $X$ case is given by
\begin{equation}
\rho(m)=\delta(m-M_X)+\frac{1}{\pi}\delta'(m).
\end{equation}
where the S-wave phase-shift as a function of the invariant mass
vanishes below the $D \bar D^*$ threshold. For the unbound case, the
bound state contribution $\delta(m-M_X)$ is simply dropped out. Note
that from Fig.~\ref{fig1} the phase-shift at low energies is a
decreasing function, so its derivative becomes negative which is the
essence of the Dashen-Kane cancellation. If the mass of $X$ is not
correctly reconstructed, because we have a finite resolution in our
detector, given by the response function $R_\sigma(m,M)$, we will measure real $DD^*$ pairs from the decay of the $X$ and $DD^*$
from the continuum, so that we cannot distinguish them due to the
finite detector resolution. Thus, we have to fold the detector
response function and the density of states as done in Eq.~\eqref{eq:smearedrho}
applied to the $X(3872)$ case 
\begin{eqnarray}
 \bar{\rho}_\sigma(M) = \Theta\,R_\sigma(M_X,M)+\frac{1}{\pi}\int_{M_{DD^*}}^{\infty} R_\sigma (m,M)\delta'(m) dm  
\end{eqnarray}
being $M_{DD^*}$ the $DD^*$ threshold mass and
$\Theta\equiv\Theta(M_{DD^*}-M_X)$ the Heaviside function. We show in
Fig.~\ref{fig2} the smear of the density of states for $E_b=180$ keV (bound)
and $E_b=-180$ keV (virtual) for different resolutions in the range
$\sigma=1-6$MeV. When $E_b \gg \sigma$ the finite resolution does not
modify the lineshape and effectively corresponds to $\sigma \to 0$
picture. For finite $\sigma$ the cancellation becomes rather evident
and is more effective for larger resolutions $\sigma \gg |E_b|$ where
the difference between a bound and a virtual state becomes small.

\subsection{Missing decays vs missing counts}

According to a recent work, there are a number (about a third) of
unknown decays when absolute branching ratios are considered and
compared to the total width of the $X(3872)$~\cite{Li:2019kpj} (see
also \cite{Lees:2019xea} for an experimental upgrade) suggesting new
experiments to detect these missing decays. The statistical analysis
carried out by the authors of Ref.~\cite{Li:2019kpj} provides large
error bars for the branching $Br (X(3872) \to {\rm unknown})= 1-
\sum_i{\Gamma_i}/\Gamma=31.9^{+18.1}_{-31.5}\,\%$ from the analysis of 8
detected channels (see their table II). Actually, about half of the
decays goes into $D \bar D^*$ pairs. We note here that the quenching
effect we unveil here may be behind such missing decays, since quite
generally and due to the Dashen-Kane cancellation the counted signals
are suppressed against the original ones, $\overline{N}_{1^{++}} <
N_{X(3872)}$.  This undercounting is in complete agreement with our
previous study~\cite{Ortega:2017hpw,Ortega:2017shf} on occupation
numbers at finite temperature and of relevance in $X(3872)$ in heavy
ion-collisions. It also complies with the similarities of production
rates at finite $p_T$ of deuterons and $X(3872)$ states in pp
collisions at ultrahigh energies in the mid-rapidity
region~\cite{Ortega:2019fme} which provides, after correcting the effect
to a one-to-one production rate, $N_X/N_d \sim 1$.

\begin{figure*}[ht]
  \includegraphics[width=.45\textwidth]{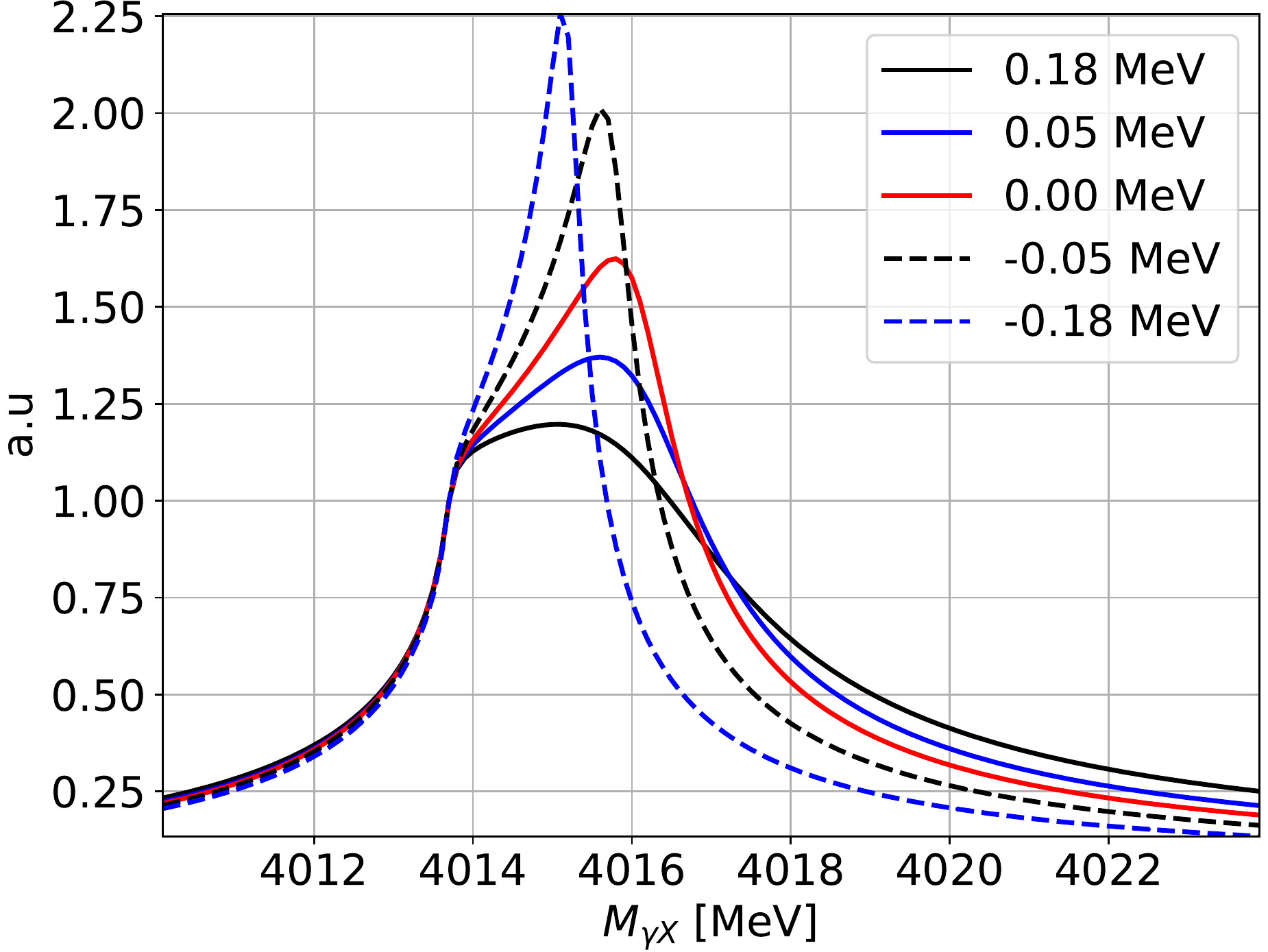}
  \includegraphics[width=.45\textwidth]{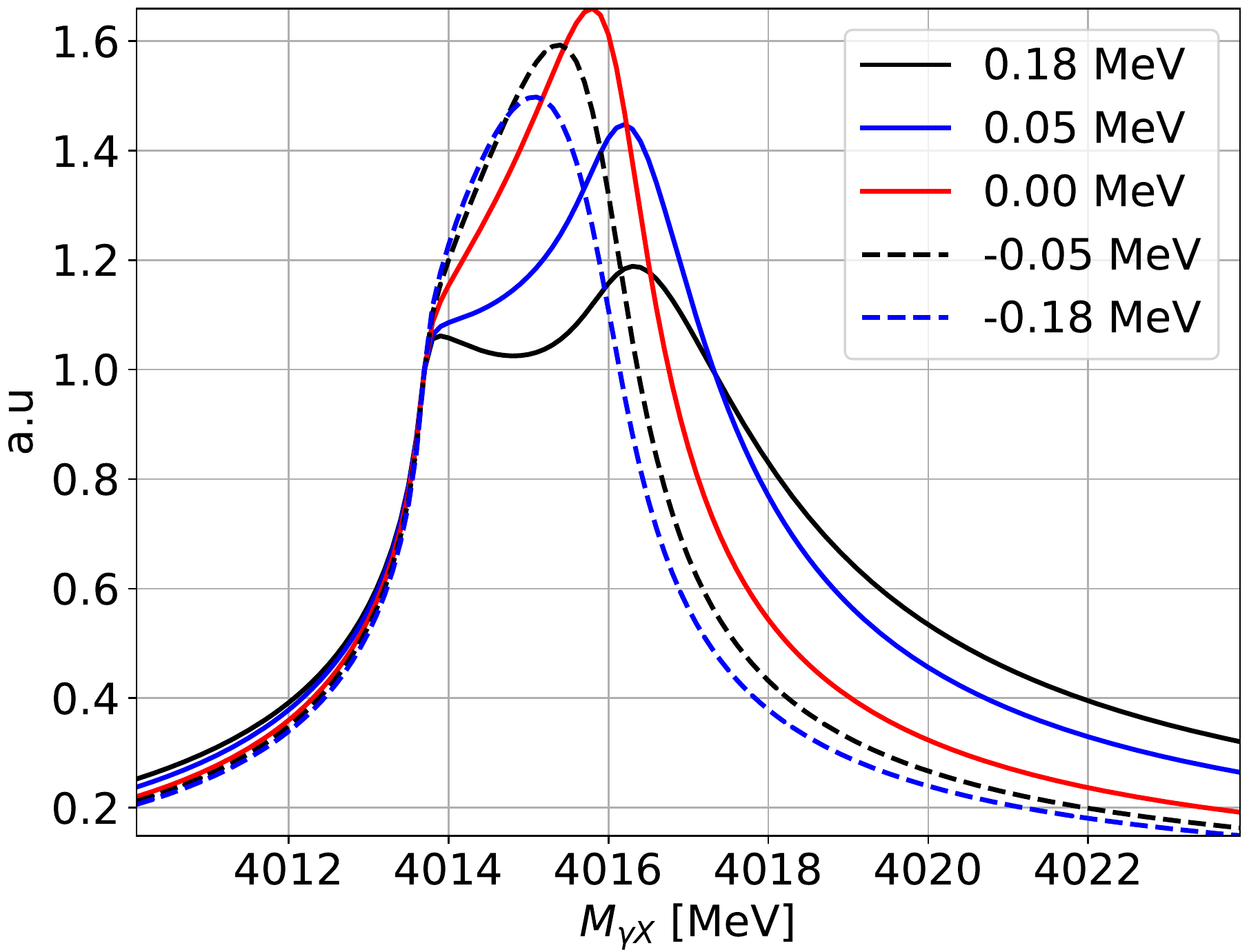}
  \includegraphics[width=.45\textwidth]{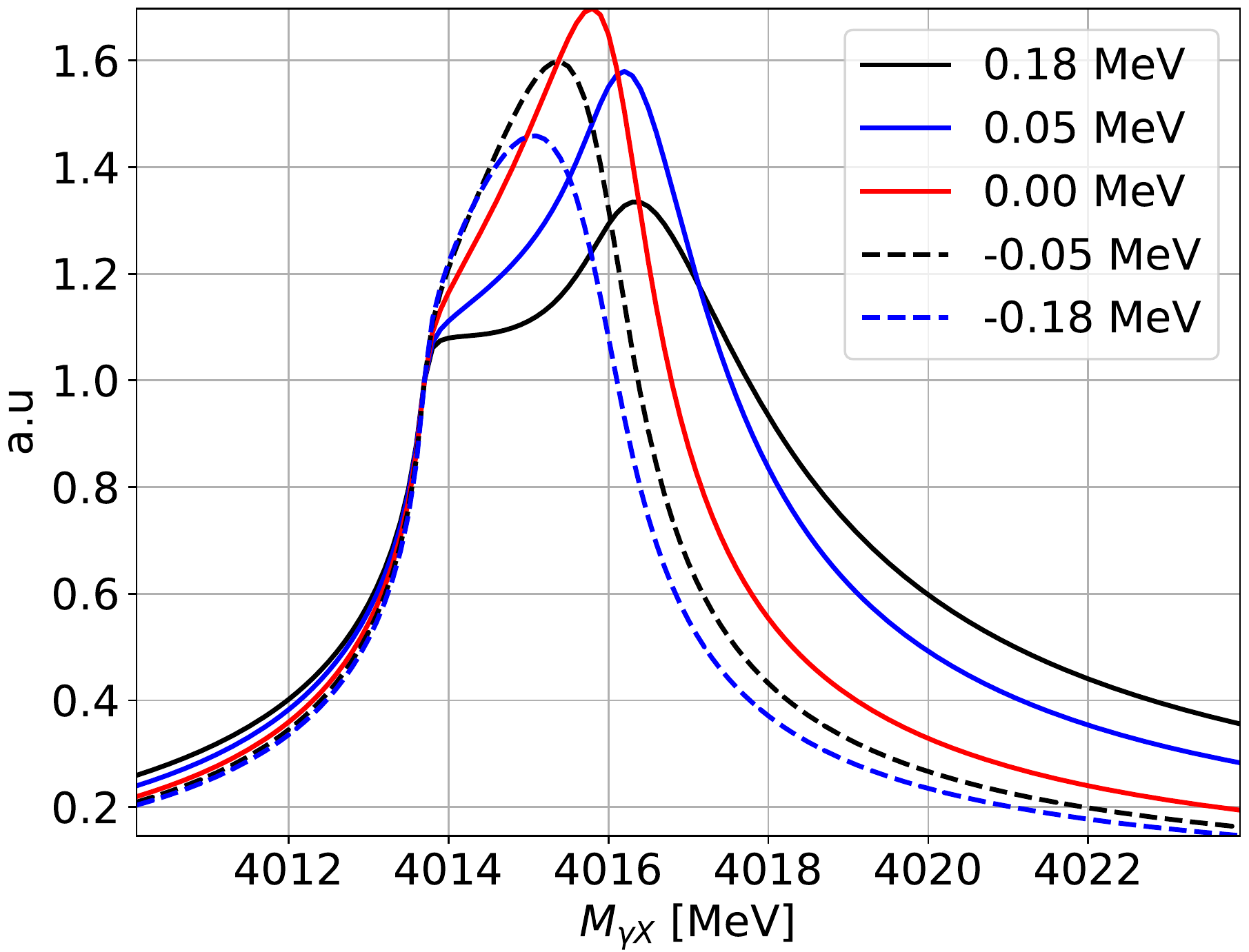}
  \includegraphics[width=.45\textwidth]{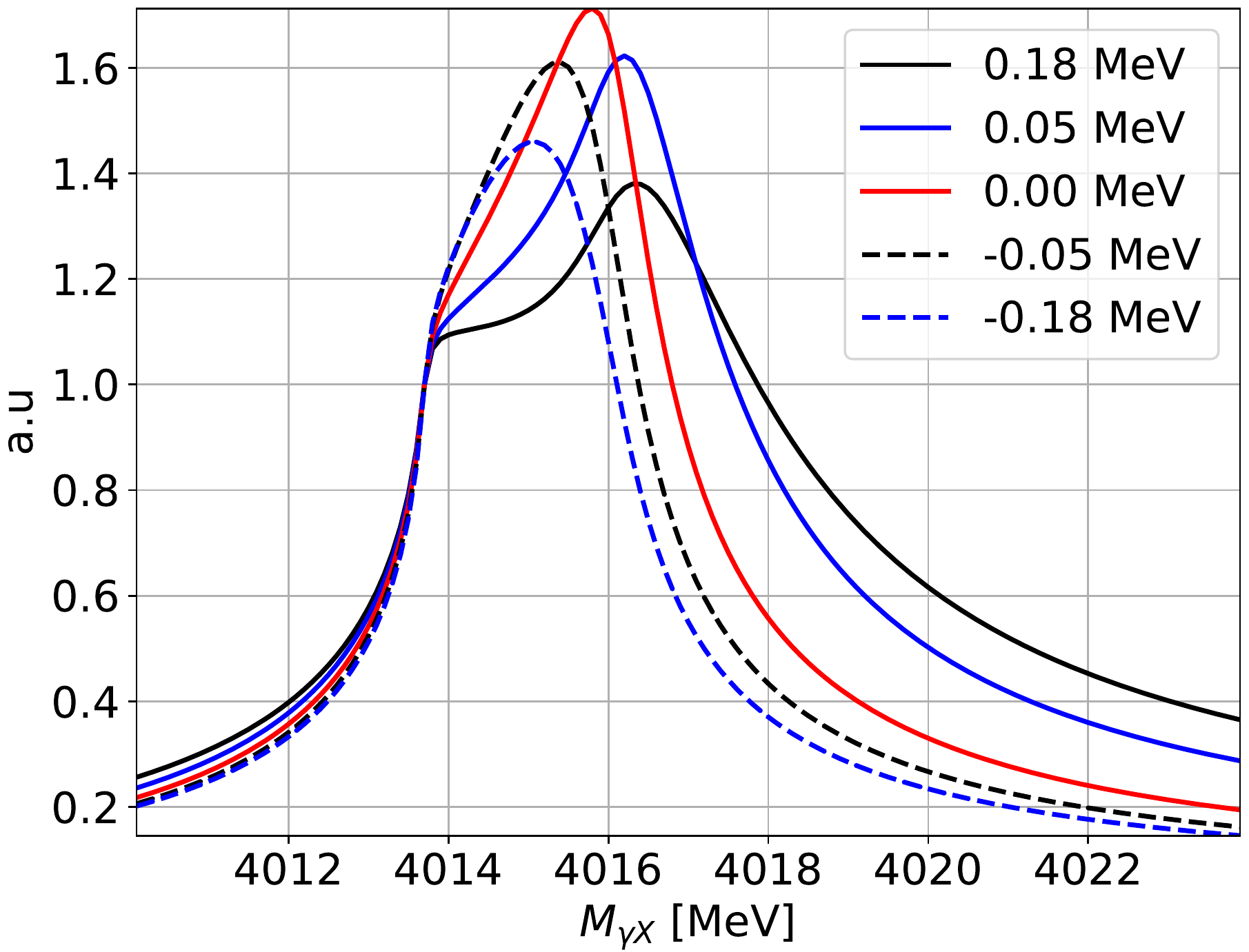}
  \caption{\label{fig4} Smeared lineshapes of states, $\bar{L}(s)$,
    for $\sigma=0$ MeV (top,left), $\sigma=1$ MeV (top,right) ,
    $\sigma=3$ MeV (bottom,left), and $\sigma=4$ MeV (bottom,right)
    for the S-wave source of Ref.~\cite{Guo:2019qcn}, for different
    binding energies and a $\Delta m=2$ MeV.}
\end{figure*}

\begin{figure*}[ht]
  \includegraphics[width=.45\textwidth]{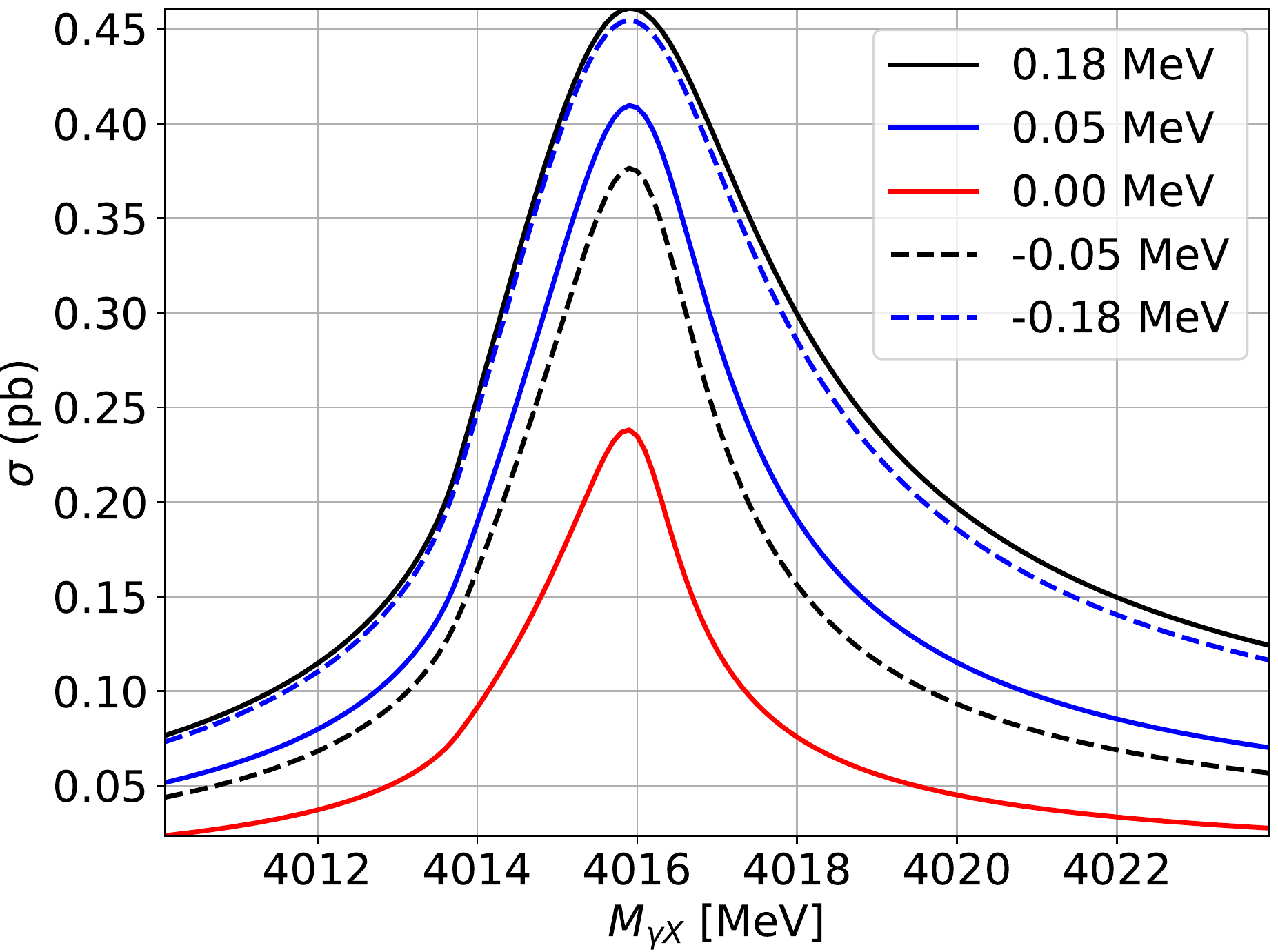}
  \includegraphics[width=.45\textwidth]{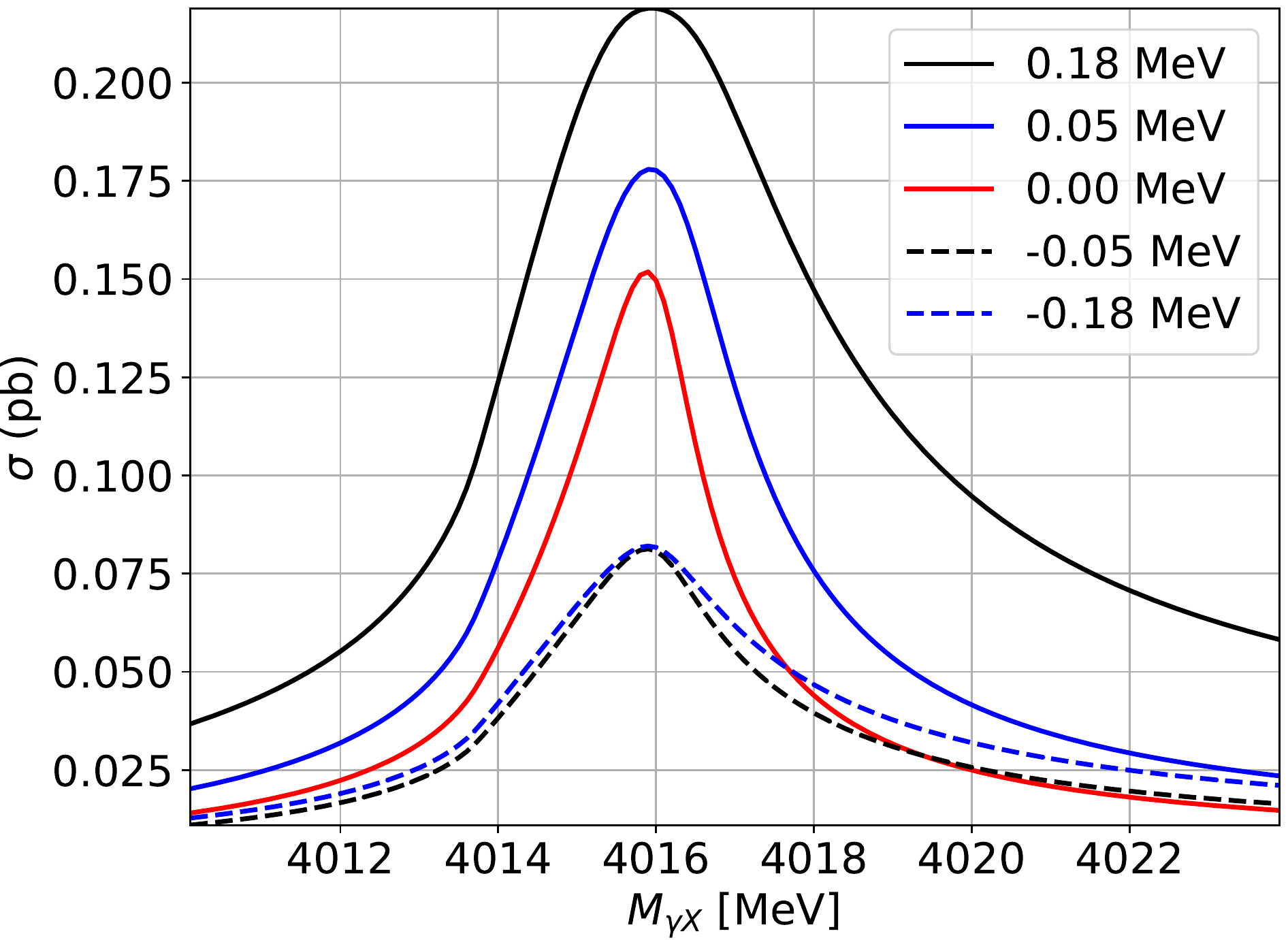}
  \includegraphics[width=.45\textwidth]{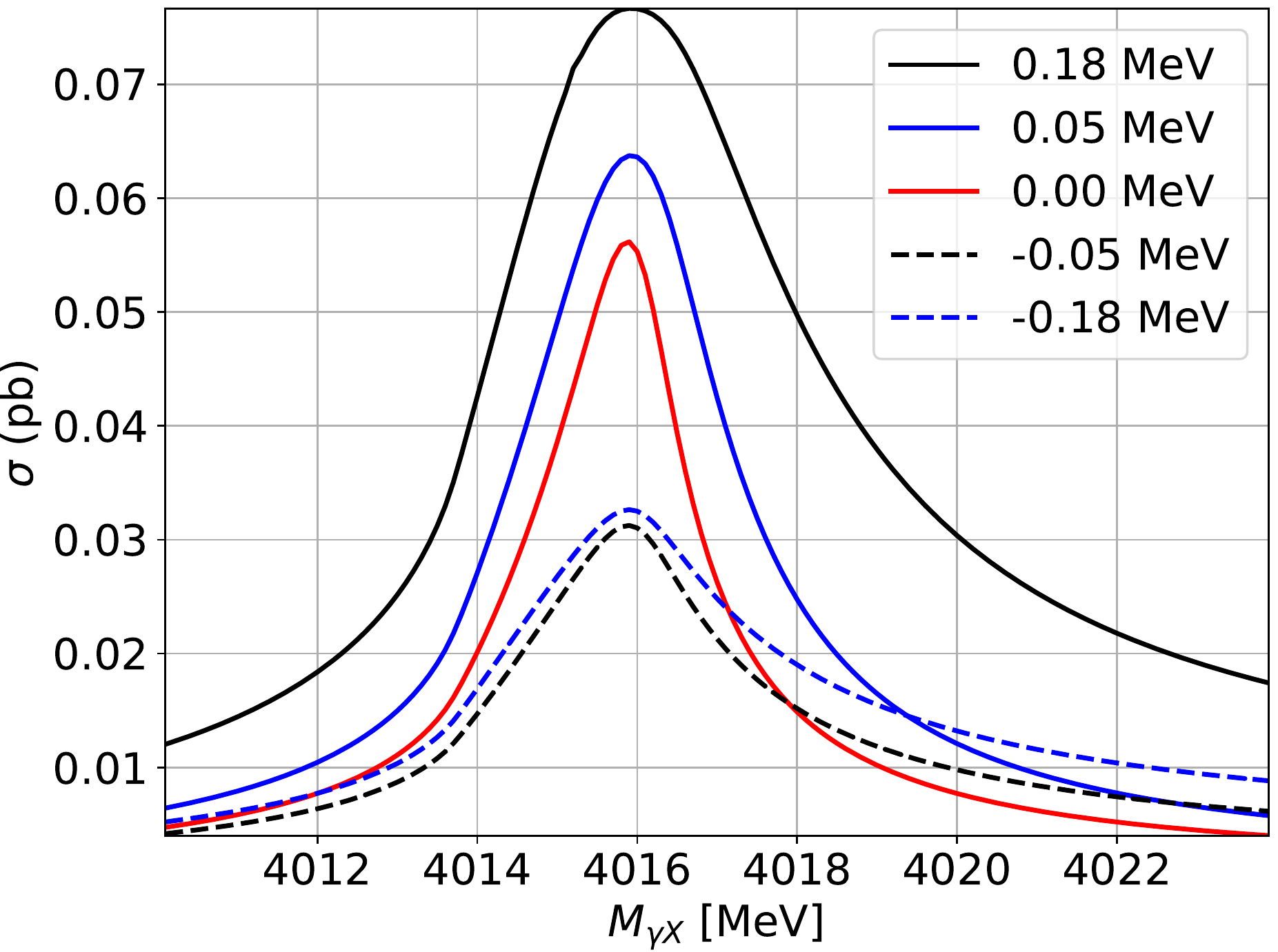}
  \includegraphics[width=.45\textwidth]{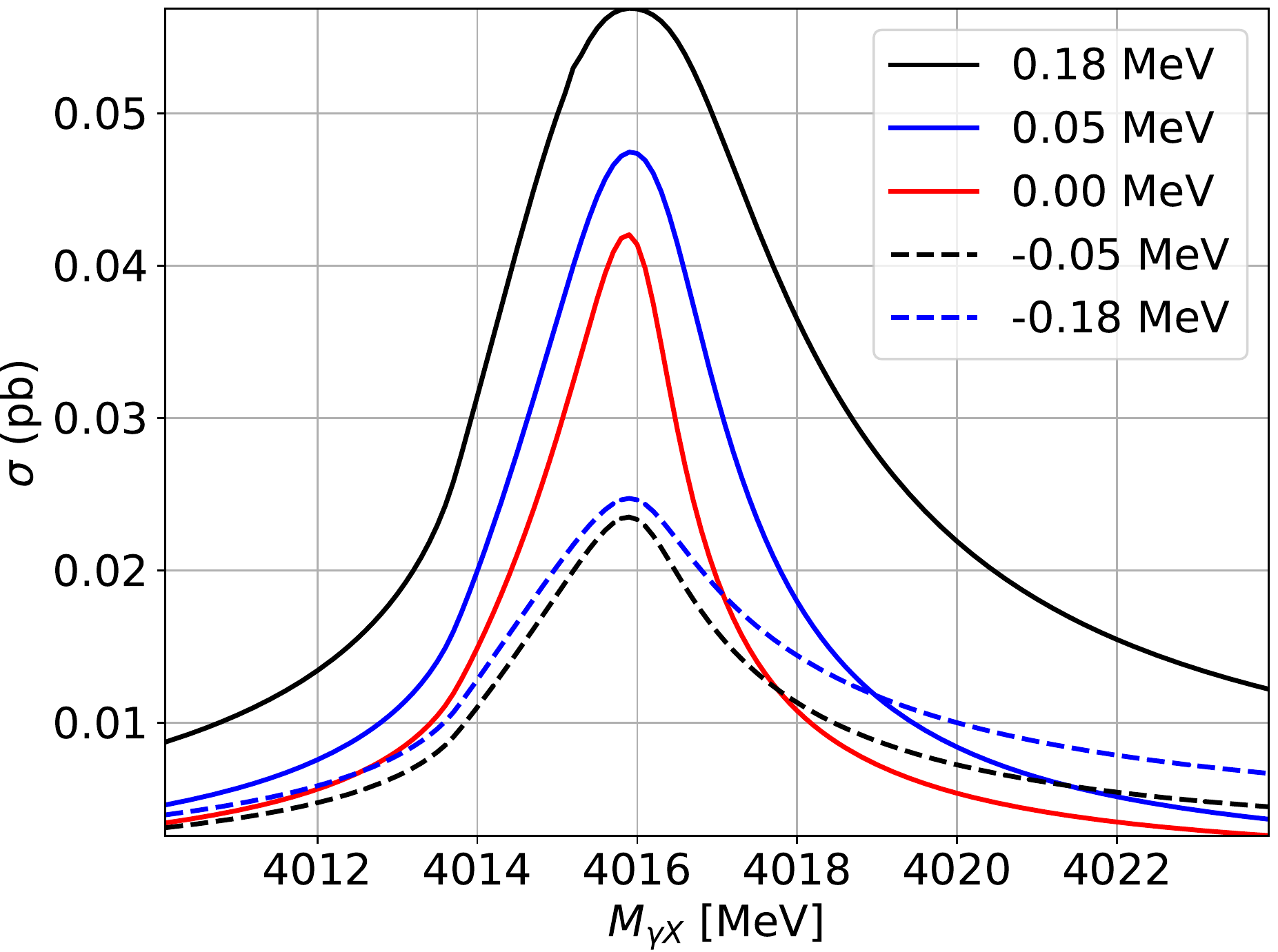}
  \caption{\label{fig6} Smeared lineshapes of states, $\bar{L}(s)$,
    for $\sigma=0$ MeV (top,left), $\sigma=1$ MeV (top,right) ,
    $\sigma=3$ MeV (bottom,left), and $\sigma=4$ MeV (bottom,right)
    for the P-wave source of Ref.~\cite{Braaten:2019gfj}, for
    different binding energies and a $\Delta m=2$ MeV.}
\end{figure*}

\section{Smearing of lineshapes}
\label{sec:num}

\subsection{General considerations}

As we have discussed above, the finite detector resolution does not
separate between the signals triggered by a bound $X(3872)$ and $D \bar
D^*$ pairs in the $1^{++}$ nearby continuum. This fact in itself
should not necessarily be a cause of concern {\it if} the level
density was a smooth function within the finite resolution
$\sigma$. However, we have seen that this is {\it not} what happens in
the $1^{++}$ channel; a relevant variation with positive and negative
contributions does take place. This, of course, sets the problem on how
would it be possible to deduce accurately the mass of the $X(3872)$
state given these limitations on resolution and being aware of the
cancellation effect.

In general, the direct determination of the mass would require more
precision on the mass of the constituents (i.e., $D^0$ and $D^{*\,0}$
mass assuming a molecular nature) and a large acquisition of
statistics, considering the small value of the $X$ binding energy. An
alternative, and more interesting method, is the characterization of
production processes in terms of a suitable mass operator ${\cal
  O}(M)$, sensitive to small variations of the binding
energy. Recently, two methods involving triangle singularities near
the $D^{*\,0}\bar D^{*\,0}$ threshold have been
proposed~\cite{Guo:2019qcn,Braaten:2019gfj}. Those kinematic
singularities, which are formed when the three particles composing the
triangle are simultaneously on-shell, have been suggested to provide a
more accurate method to determine the $X(3872)$ binding energy than
direct mass measurements.

\subsection{Smearing effects}

All experimental analysis make a
distinction between the resolution $\sigma$ of the detectors leading
to a gaussian response function for a monochromatic signal with
invariant mass $m_0$
\begin{eqnarray}
\delta ( m- m_0) \to \frac1{\sqrt{2\pi} \sigma} e^{- \frac12 \left(
  \frac{m-m_0}\sigma \right)^2}
\end{eqnarray}
This is one source of mixing mass effects. On the other hand, a choice
of mass window, $\Delta m$ for measuraments must be made.  This is
another source of mass mixing, since the resulting signal will
correspond to the averaged mass distribution around a chosen mass
interval $\pm \Delta m/2$.

In order to illustrate the aforementioned limitations due to the resolution and the cancellation effect, let's consider now a general lineshape $L(s,M)$, where $s$ is the
invariant mass and $M$ is the reconstructed mass of the secondary $X$
particle from the Gaussian distribution
  $R_\sigma(m,M)$.  The convoluted lineshape from
  the $X$ particle with mass $M$ is (Eq.~\ref{eq:Obs-can})

\begin{align}
 \bar{L}(s) &=\Theta\,{\cal R}(M_X) L(s,M_X)+\frac{1}{\pi}\int_{M_{DD^*}}^{\infty} {\cal R}(m)L(s,m)\delta'(m) dm 
\end{align}
with  ${\cal R}(m)=\frac{1}{2\Delta m}\left[{\rm Erf}\left(\frac{m-M_X+\Delta m/2}{\sqrt{2}\sigma}\right)+{\rm Erf}\left(\frac{M_X-m+\Delta m/2}{\sqrt{2}\sigma}\right)\right]$.

We analyze the effect of smearing for the lineshapes generated in the
$X(3872) \gamma $ production process using either a relative
S-wave~\cite{Guo:2019qcn} or P-wave~\cite{Braaten:2019gfj} source of a
$D^{*0} \bar D^{*0}$ pair.  Results for the S-wave source of
Ref.~\cite{Guo:2019qcn} can be see in Fig.~\ref{fig4} and results for
the P-wave source of Ref.~\cite{Braaten:2019gfj} are shown in
Fig.~\ref{fig6}, without considering a finite binning in the $\gamma
X$ invariant mass spectrum. For the S-wave source, normalized to the
$D^{*\,0}\bar D^{*\,0}$ threshold, we appreciate a change in the shape
of the distribution, which pretty much blurs the neat distinction due
to the $X$ binding energy. Still, we see a separation of the lineshape
tails which could be used for the latter purposes. The cancellation
and the finite resolution, thus, leads to a more complicated precise
measurement of the $X(3872)$ mass, specially when finite statistics
are considered (see discussion below). For the P-wave source, the main
effect is the absolute value decreases of the lineshapes, depending on
their binding energy due to the cancellation (effect that also occurs
for the S-wave source but it not appreciated due to the normalization
of the lineshapes).

\subsection{Finite samples}

It is interesting to analyze the S-wave source results from the
counting statistics point of view. We expect a convergence of all
$\gamma X$ lineshapes regardless of the $X$ binding energy. Their
tails decrease at different rates, but a limited statistics can
compromise their proper identification. Quite generally we will be
able to discern two different (smeared) signals if the number of
events fulfills
\begin{eqnarray}
\frac{\Delta {\cal O}}{\cal O} \sim \frac1{\sqrt{N}}
\end{eqnarray}

\begin{figure}[h]
 \includegraphics[width=.5\textwidth]{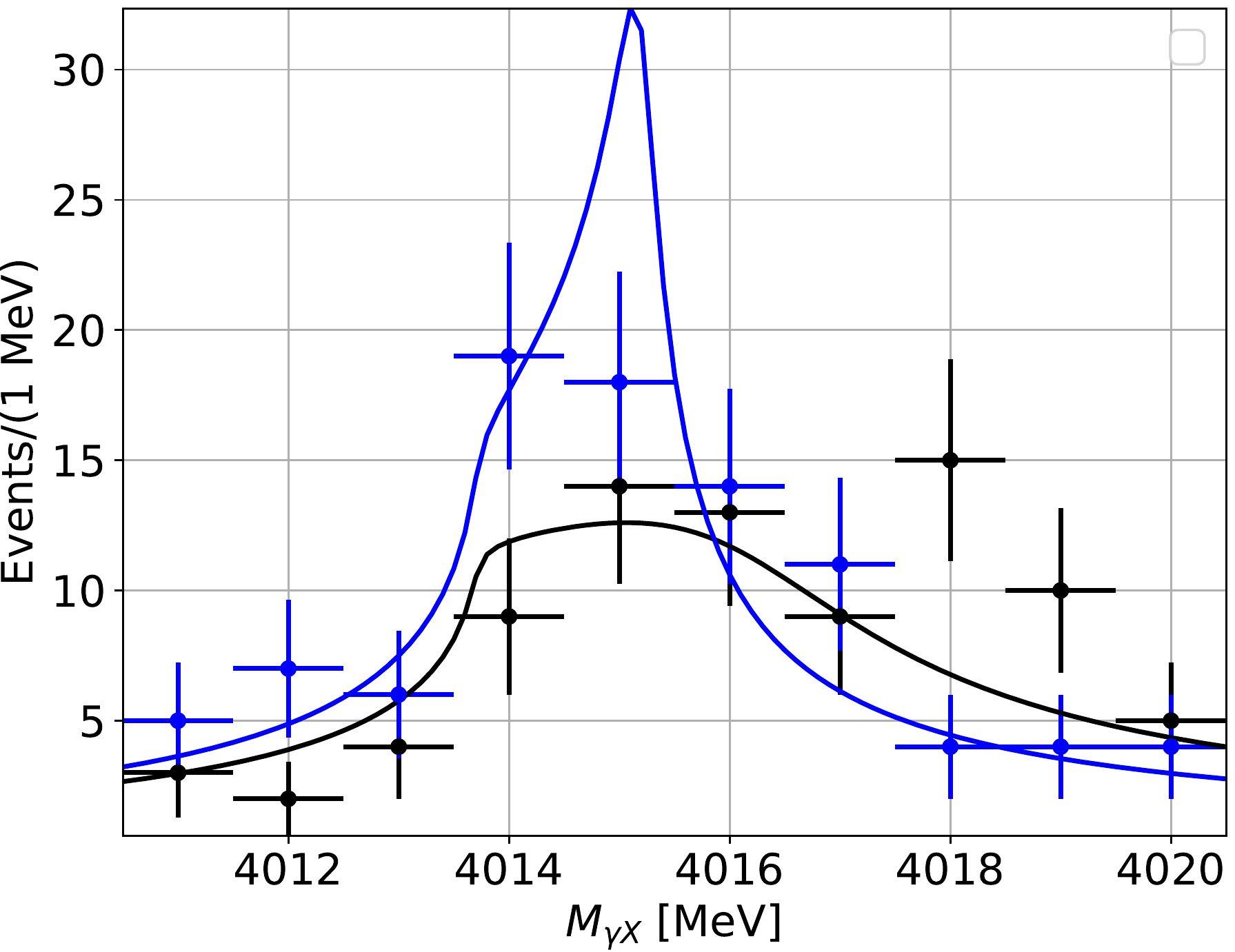}
 \includegraphics[width=.5\textwidth]{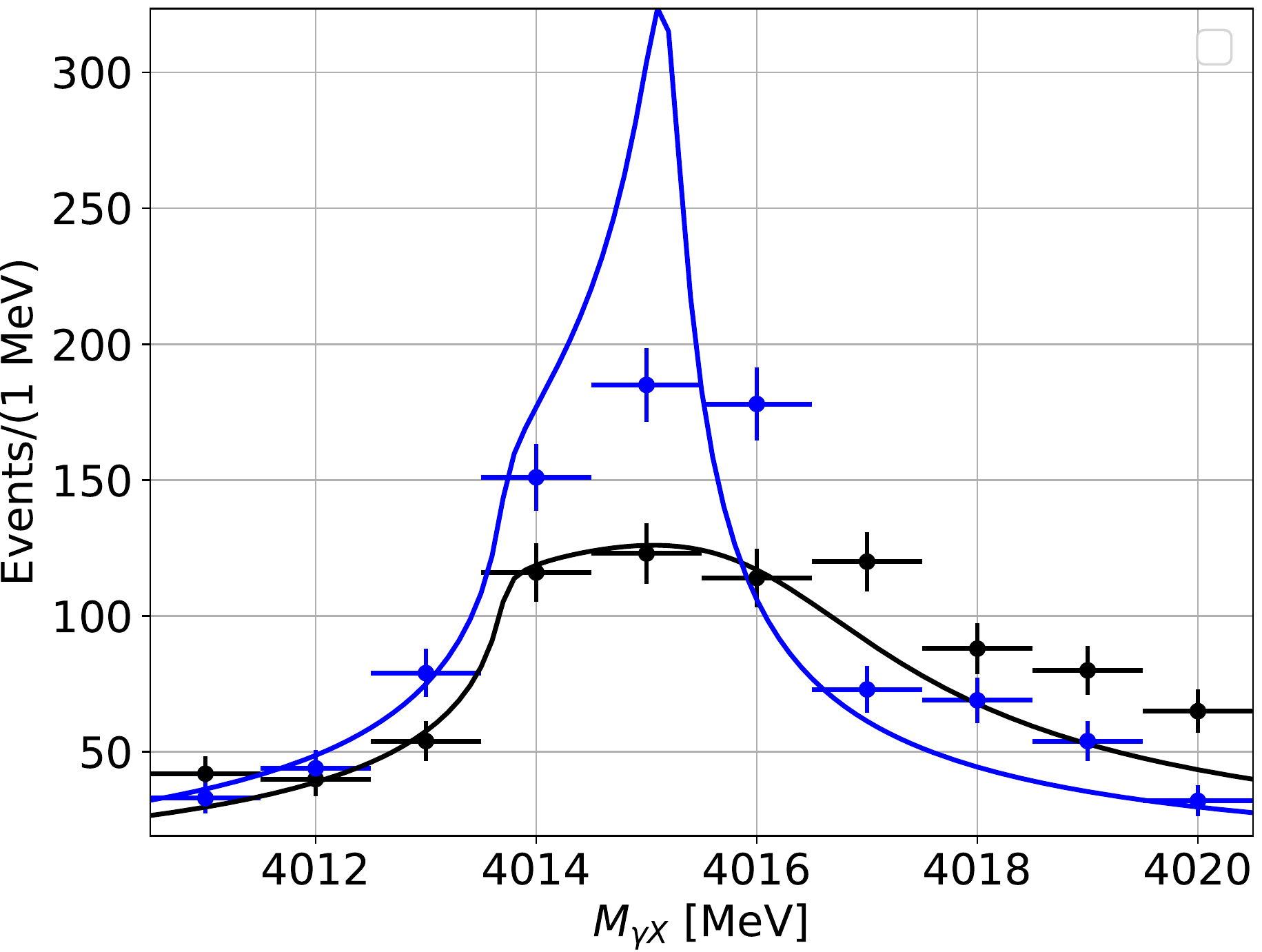}
 \caption{\label{fig5} Binned smeared lineshapes of the S-wave source for $N=100$ events (top) and $N=1000$ events (bottom). We compare the $E_b=180$ keV (black) and $E_b=-180$ keV (blue) binding energies, using a $\sigma=2$ MeV resolution, an energy bin of $1$ MeV and a $\Delta m=2$ MeV. The full lineshape
   corresponding to $\Delta M=\sigma=0$ is shown for comparison.
}
\end{figure}

In Fig.~\ref{fig5} we show an example of limited resolution for
binding energies $E_b=180$ keV and $E_b=-180$ keV (virtual), a
$\sigma=2$ MeV and an energy bin of $E_{\rm win}=1$ MeV. The synthetic
data is obtained by randomly sampling $N=100$ and $1000$ events in the
[4010,4020] energy range, according to the probability density
function given by the lineshapes of Fig.~\ref{fig4}(bottom). As the
global normalization of the lineshapes of Fig.~\ref{fig4} are not
known, the same occurs to the global normalization of the synthetic
data, so caution should be taken when direct comparing between the
lineshapes for different binding energies. Of course, for larger
values of $\sigma$ all curves resemble each other and the strong mass
dependence is largely washed out. We believe these effects should be
considered in an eventual benchmark experimental determination of the
$X(3872)$ mass.

\begin{figure}[h]
 \includegraphics[width=.5\textwidth]{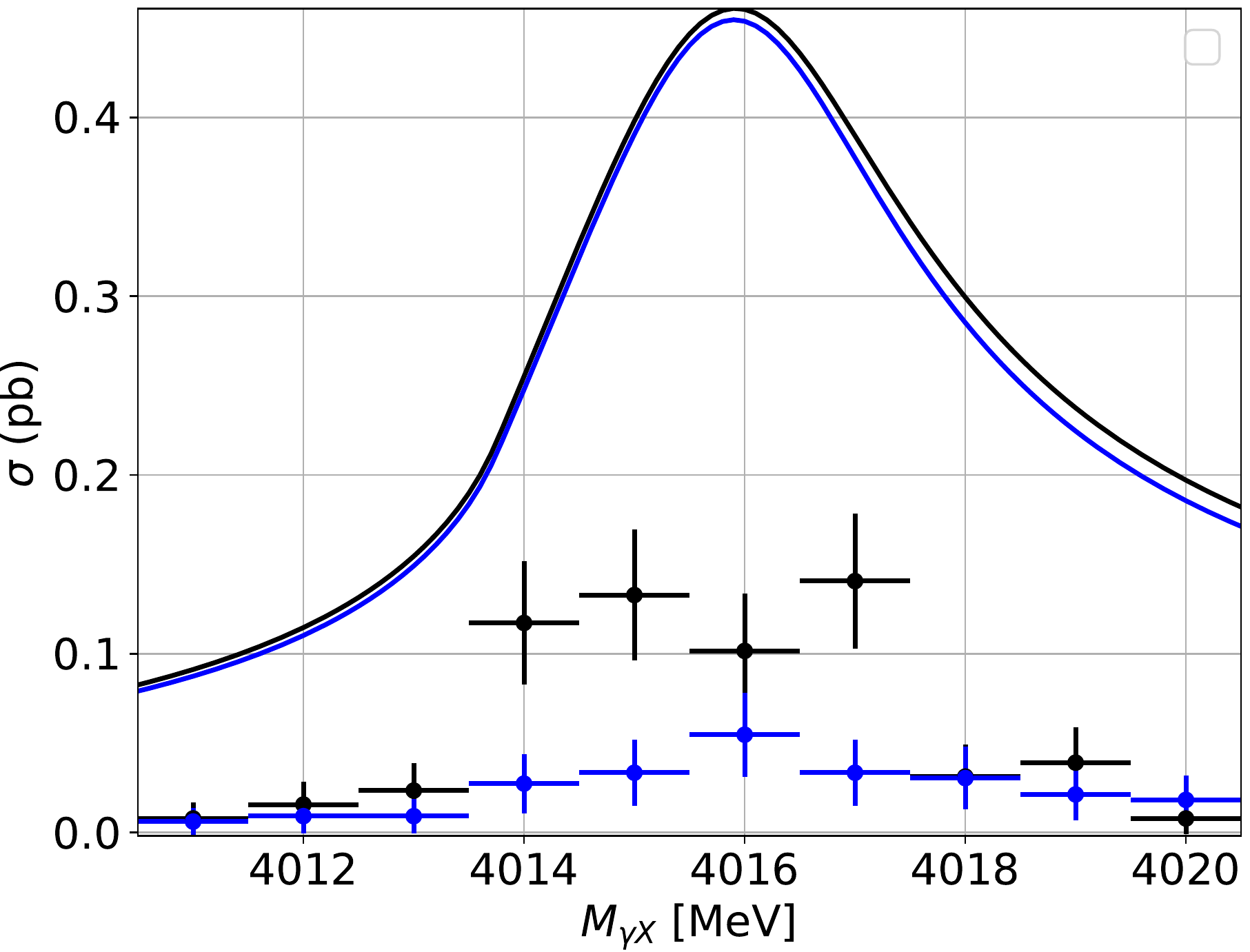}
 \includegraphics[width=.5\textwidth]{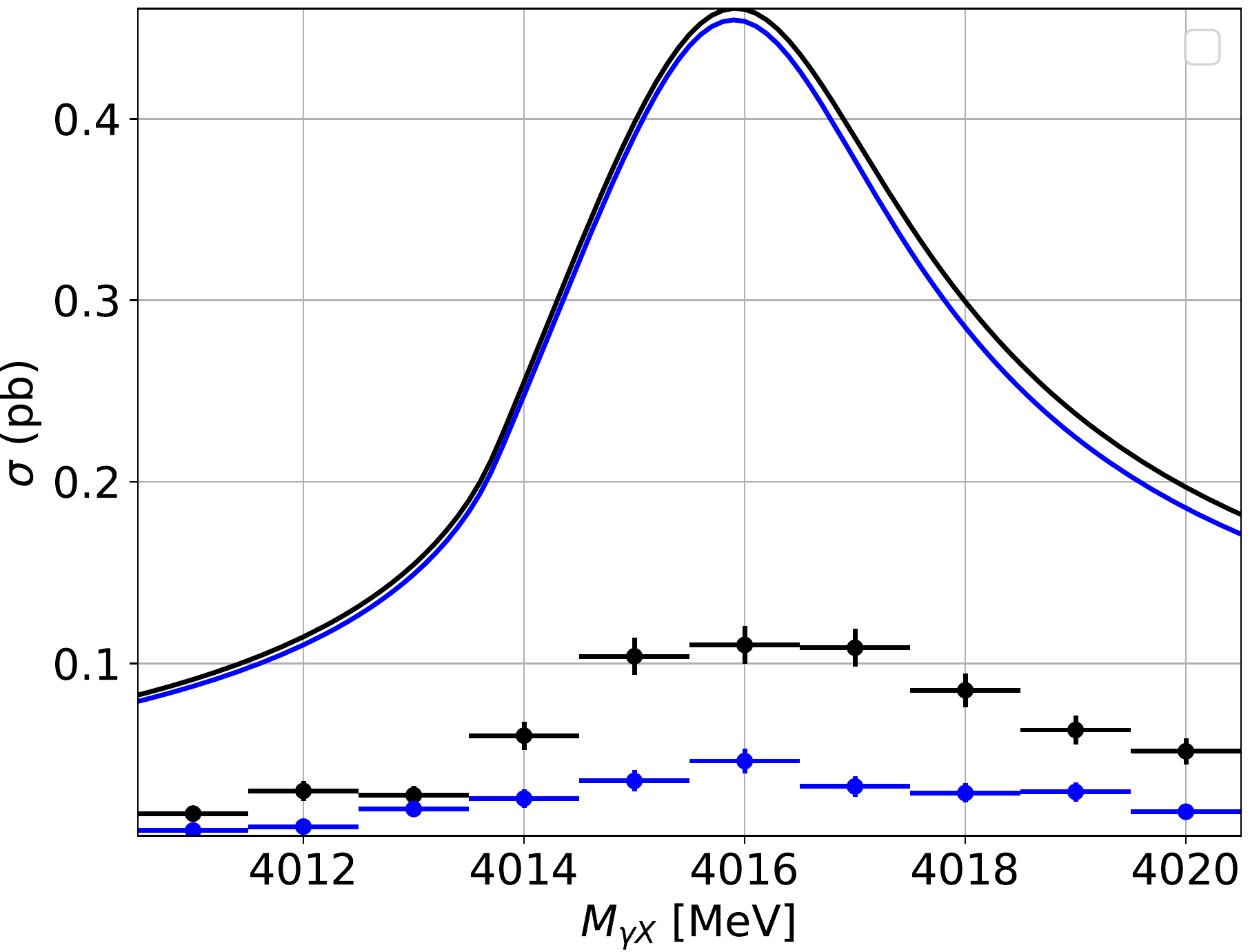}
 \caption{\label{fig7} Binned smeared lineshapes of the P-wave source for $N=100$ events (top) and $N=1000$ events (bottom). Same legend as in Fig.~\ref{fig5}.}
\end{figure}

\section{Binding independent short distance $\bar D D^* $ correlations }

\label{sec:corr}

\begin{figure}[h]
 \includegraphics[width=.5\textwidth]{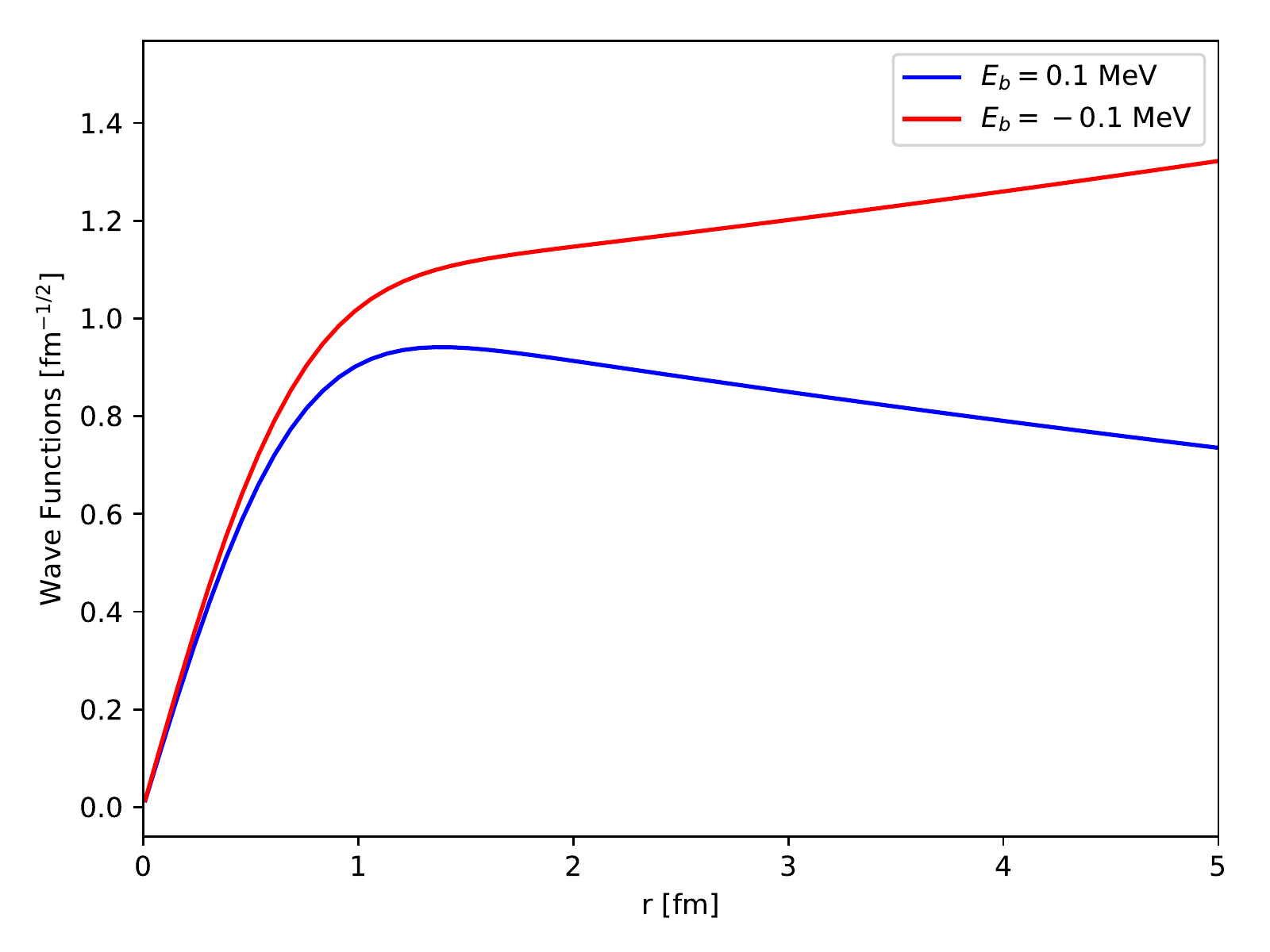}
 \caption{\label{fig:wave} Reduced $\bar D D^*$ S-wave function for
   the $1^{++}$ channel in the quark model of Ref.~\cite{Ortega:2009hj,Ortega:2012rs}. We compare
   for the bound case with $E_b=180$ keV (blue) and the unbound case
   with $E_b=-180$ keV (red) binding energies as a function of the
   relative $\bar D D^*$ distance.}
\end{figure}

  One important aspect within the present context is that, regardless
  of the precise features of the lineshape, the existence of the peak
  does not depend crucially on the $X(3872)$ being truely bound or
  unbound. At long distances the reduced relative wave functions
  for a bound/unbound $X(3872)$ state behaves as
  \begin{eqnarray}
  u_X (r) \to A_X e^{\mp \gamma_X r}   
  \end{eqnarray}
  respectively where $\gamma_X$ is the corresponding wave number which
  corresponds to a pole of the partial wave amplitude.  Clearly, since
  the signal for the $X(3872)$ is reconstructed by detecting its weak decay products,
  which involves a short distance operator and tells us about the
  relative $\bar D D^*$ wave function at {\it short distances}. To further
  analyze this issue, we plot in Fig.~\ref{fig:wave} the case of bound/unbound
wave functions, normalized so that their long-distance
  extrapolated value to the origin is unity, i.e. $ u_X (r) \to e^{\mp
    \gamma_X r} $, using for illustration the particular (unquenched)
  quark model of Ref.~\cite{Ortega:2009hj,Ortega:2012rs} which
  includes both a $c \bar c$ and $D \bar D^*$ channels in the
  description of the $X(3872)$ state.  As we see, their short distance
  behaviour is nearly identical although they are completely different
  at long distances~\footnote{Remarkably, these features between bound
    or unbound $X(3872)$ are also present in the neutron-proton case;
    while in the triplet $^3S_1$ channel they form a bound state, in
    the singlet $^1S_0$ channel they are unbound. Nevertheless, their
    wave functions are rather similar at short distances (see e.g.
    \cite{CalleCordon:2008cz} for a discussion).}. Thus, the bumps
  found experimentally and attributed to be the $X(3872)$ are
  certainly reflecting a strong short distance correlated $\bar D D^*$
  pair in the $1^{++}$ channel, and not just a bound state
  feature. The short distance dominance of the $X(3872)$ is not new,
  and it provides an explanation for the large isoscalar to isovector
  decay modes (see e.g. ~\cite{Gamermann:2009uq}).

\section{Conclusions}
\label{sec:concl}

In this paper, we have analyzed the impact of finite detector
resolution in the production and decay of the $X(3872)$ state. We have
discussed the cancellation effect due to the superposition in the
level density in the $1^{++}$ channel of bound state and nearby $D
\bar D^*$ continuum states in the initial state, which cannot be
separated in the final state when the binding energy is much smaller
than the energy resolution.  Our results suggest that the mechanism of
production of weakly bound states such as the $X(3872)$ undercounts
the number of states $\overline{N}_{1^{++}} < N_{X(3872)}$, an effect
which is in harmony with the missing resonances reported in a recent
absolute branching ratio analysis. This signal suppression is in
complete agreement with our previous study on occupation numbers at
finite temperature and of relevance in $X(3872)$ in heavy
ion-collisions. It also complies with the deuteron to X(3872) finite
$p_T$ production ratio in pp collisions at ultrahigh energies at
mid-rapidity. Our findings are also relevant to future benchmark
determinations of the $X(3872)$, particularly those displayed by the
strong lineshape dependence in production processes involving triangle
singularities. Quite generally we find that the initial density of
states triggering a signal of $X$-production in a finite resolution
energy detector blurs the spectrum and hence the strong mass
dependence is reduced and could only be pinned down with sufficiently
high statistics. This is in harmony with the relevance of
short distance  $D \bar D^*$ correlations in the $1^{++}$ channel.
We expect our observations to hold in similar weakly
bound states not directly measured through their track, but inferred
from their decay products.

\bibliographystyle{h-elsevier}
\bibliography{Xline,references}

\begin{thebibliography}{10}

\bibitem{PDG}
P.Z. et~al. (Particle Data~Group),
\newblock Prog. Theor. Exp. Phys.  (2020) 083C01.

\bibitem{RuizArriola:2016qpb}
E. Ruiz~Arriola et~al.,
\newblock {Excited Hyperons in QCD Thermodynamics at Freeze-Out}, pp. 128--139,
  2016, 1612.07091.

\bibitem{Choi:2003ue}
Belle, S.K. Choi et~al.,
\newblock Phys. Rev. Lett. 91 (2003) 262001, hep-ex/0309032.

\bibitem{Aubert:2004zr}
BaBar, B. Aubert et~al.,
\newblock Phys. Rev. D71 (2005) 031501, hep-ex/0412051.

\bibitem{Choi:2011fc}
Belle, S.K. Choi et~al.,
\newblock Phys. Rev. D 84 (2011) 052004, 1107.0163.

\bibitem{Aaij:2013zoa}
LHCb, R. Aaij et~al.,
\newblock Phys. Rev. Lett. 110 (2013) 222001, 1302.6269.

\bibitem{Godfrey:2008nc}
S. Godfrey and S.L. Olsen,
\newblock Ann. Rev. Nucl. Part. Sci. 58 (2008) 51, 0801.3867.

\bibitem{Guo:2017jvc}
F.K. Guo et~al.,
\newblock Rev. Mod. Phys. 90 (2018) 015004, 1705.00141.

\bibitem{Brambilla:2019esw}
N. Brambilla et~al.,
\newblock (2019), 1907.07583.

\bibitem{Aaij:2020xjx}
LHCb, R. Aaij et~al.,
\newblock JHEP 08 (2020) 123, 2005.13422.

\bibitem{Guo:2019qcn}
F.K. Guo,
\newblock Phys. Rev. Lett. 122 (2019) 202002, 1902.11221.

\bibitem{Braaten:2019gfj}
E. Braaten, L.P. He and K. Ingles,
\newblock Phys. Rev. D100 (2019) 031501, 1904.12915.

\bibitem{Dashen:1974ns}
R.F. Dashen and G.L. Kane,
\newblock Phys. Rev. D11 (1975) 136.

\bibitem{Ortega:2017hpw}
P.G. Ortega et~al.,
\newblock Phys. Lett. B781 (2018) 678, 1707.01915.

\bibitem{Ortega:2017shf}
P.G. Ortega and E. Ruiz~Arriola,
\newblock PoS Hadron2017 (2018) 236, 1711.10193.

\bibitem{Ortega:2019fme}
P.G. Ortega and E. Ruiz~Arriola,
\newblock Chinese Physics C 43 (2019) 124107, 1907.01441.

\bibitem{RuizArriola:2020ijw}
E. Ruiz~Arriola and P.G. Ortega,
\newblock {12th International Winter Workshop "Excited QCD" 2020}, 2020,
  2005.01531.

\bibitem{Beth:1937zz}
E. Beth and G. Uhlenbeck,
\newblock Physica 4 (1937) 915.

\bibitem{Dashen:1969ep}
R. Dashen, S.K. Ma and H.J. Bernstein,
\newblock Phys. Rev. 187 (1969) 345.

\bibitem{Dashen:1974jw}
R.F. Dashen and R. Rajaraman,
\newblock Phys. Rev. D10 (1974) 694.

\bibitem{Dashen:1974yy}
R.F. Dashen and R. Rajaraman,
\newblock Phys. Rev. D10 (1974) 708.

\bibitem{Lo:2017sde}
P.M. Lo,
\newblock Eur. Phys. J. C 77 (2017) 533, 1707.04490.

\bibitem{Lo:2020phg}
P.M. Lo,
\newblock (2020), 2007.03392.

\bibitem{Fukuda:1956zz}
N. Fukuda and R. Newton,
\newblock Phys. Rev. 103 (1956) 1558.

\bibitem{DeWitt:1956be}
B.S. DeWitt,
\newblock Phys. Rev. 103 (1956) 1565.

\bibitem{Gomez-Rocha:2019xum}
M. Gómez-Rocha and E. Ruiz~Arriola,
\newblock Phys. Lett. B 800 (2020) 135107, 1910.10560.

\bibitem{Gomez-Rocha:2019rpj}
M. Gómez-Rocha and E. Ruiz~Arriola,
\newblock Phys. Rev. D 101 (2020) 036003, 1911.08990.

\bibitem{Dashen:1976cf}
R.F. Dashen, J.B. Healy and I.J. Muzinich,
\newblock Phys. Rev. D14 (1976) 2773.

\bibitem{knoll2010radiation}
G.F. Knoll,
\newblock Radiation detection and measurement (John Wiley \& Sons, 2010).

\bibitem{Arriola:2015gra}
E. Ruiz~Arriola, L.L. Salcedo and E. Megias,
\newblock Acta Phys. Polon. Supp. 8 (2015) 439, hep-ph/1505.02922.

\bibitem{Arriola:2014bfa}
E. Ruiz~Arriola, L.L. Salcedo and E. Megias,
\newblock Acta Phys. Polon. B45 (2014) 2407, hep-ph/1410.3869.

\bibitem{Horowitz:2005nd}
C.J. Horowitz and A. Schwenk,
\newblock Nucl. Phys. A776 (2006) 55, nucl-th/0507033.

\bibitem{Nussinov:1976fg}
S. Nussinov and D.P. Sidhu,
\newblock Nuovo Cim. A44 (1978) 230.

\bibitem{Tanabashi:2018oca}
Particle Data Group, M. Tanabashi et~al.,
\newblock Phys. Rev. D98 (2018) 030001.

\bibitem{Esposito:2016noz}
A. Esposito, A. Pilloni and A.D. Polosa,
\newblock Phys. Rept. 668 (2016) 1, 1611.07920.

\bibitem{Karliner:2017qhf}
M. Karliner, J.L. Rosner and T. Skwarnicki,
\newblock (2017), 1711.10626.

\bibitem{Kang:2016jxw}
X.W. Kang and J.A. Oller,
\newblock Eur. Phys. J. C77 (2017) 399, 1612.08420.

\bibitem{Braaten:2007dw}
E. Braaten and M. Lu,
\newblock Phys. Rev. D76 (2007) 094028, 0709.2697.

\bibitem{Karplus:1958zz}
R. Karplus, C.M. Sommerfield and E.H. Wichmann,
\newblock Phys. Rev. 111 (1958) 1187.

\bibitem{Szczepaniak:2015eza}
A.P. Szczepaniak,
\newblock Phys. Lett. B747 (2015) 410, 1501.01691.

\bibitem{Oset:2018bjl}
E. Oset et~al.,
\newblock Few Body Syst. 59 (2018) 85.

\bibitem{Liu:2015taa}
X.H. Liu, M. Oka and Q. Zhao,
\newblock Phys. Lett. B753 (2016) 297, 1507.01674.

\bibitem{Sakai:2020ucu}
S. Sakai, E. Oset and F.K. Guo,
\newblock Phys. Rev. D 101 (2020) 054030, 2002.03160.

\bibitem{Molina:2020kyu}
R. Molina and E. Oset,
\newblock Eur. Phys. J. C 80 (2020) 451, 2002.12821.

\bibitem{Tornqvist:1993ng}
N.A. Tornqvist,
\newblock Z. Phys. C61 (1994) 525, hep-ph/9310247.

\bibitem{Close:2003sg}
F.E. Close and P.R. Page,
\newblock Phys. Lett. B578 (2004) 119, hep-ph/0309253.

\bibitem{Braaten:2003he}
E. Braaten and M. Kusunoki,
\newblock Phys. Rev. D69 (2004) 074005, hep-ph/0311147.

\bibitem{Perez:2013jpa}
R. Navarro~Pérez, J. Amaro and E. Ruiz~Arriola,
\newblock Phys. Rev. C 88 (2013) 064002, 1310.2536,
\newblock [Erratum: Phys.Rev.C 91, 029901 (2015)].

\bibitem{Ortega:2009hj}
P.G. Ortega et~al.,
\newblock Phys. Rev. D81 (2010) 054023, hep-ph/0907.3997.

\bibitem{Ortega:2012rs}
P.G. Ortega, D.R. Entem and F. Fernandez,
\newblock J. Phys. G40 (2013) 065107, hep-ph/1205.1699.

\bibitem{Cincioglu:2016fkm}
E. Cincioglu et~al.,
\newblock Eur. Phys. J. C76 (2016) 576, 1606.03239.

\bibitem{Gamermann:2009uq}
D. Gamermann et~al.,
\newblock Phys. Rev. D81 (2010) 014029, hep-ph/0911.4407.

\bibitem{Arriola:2013era}
E. Ruiz~Arriola, S. Szpigel and V. Timoteo,
\newblock Phys. Lett. B 728 (2014) 596, 1307.1231.

\bibitem{Ablikim:2013dyn}
BESIII, M. Ablikim et~al.,
\newblock Phys. Rev. Lett. 112 (2014) 092001, 1310.4101.

\bibitem{Aaij:2011sn}
LHCb, R. Aaij et~al.,
\newblock Eur. Phys. J. C 72 (2012) 1972, 1112.5310.

\bibitem{Li:2019kpj}
C. Li and C.Z. Yuan,
\newblock Phys. Rev. D 100 (2019) 094003, 1907.09149.

\bibitem{Lees:2019xea}
BaBar, J. Lees et~al.,
\newblock Phys. Rev. Lett. 124 (2020) 152001, 1911.11740.

\bibitem{CalleCordon:2008cz}
A. Calle~Cordon and E. Ruiz~Arriola,
\newblock Phys. Rev. C 78 (2008) 054002, 0807.2918.

\end{thebibliography}

\newpage

\appendix

\section{Details on the choice of resolution}\label{sec:appendix}

In this section we come to justify the numbers
provided in Table~\ref{tab:1}. 

\begin{itemize}

\item In Ref.~\cite{Choi:2011fc}, the authors make studies of the $\psi' \to
  \pi^+ \pi^- J/\psi$ as control sample using $ 3.635 {\rm GeV} \le M (\pi^+
  \pi^- J/\psi) \le 3.735 {\rm GeV}$  and, for X(3872) studies, they use
  $ 3.77  {\rm GeV} \le M (\pi^+ \pi^- J/\psi) \le 3.97 {\rm GeV} $.
  They select events in the range $|M(\pi^+ \pi^- J/\psi)-M_{\rm peak}|\le 0.009$ GeV, taking
  energy bins of $\Delta m=2$ MeV.
  The $M(\pi^+ \pi^- J/\psi)$ mass resolution of the Belle detector in the mass region of the X(3872)
  is claimed to be $\sigma \simeq 4$ MeV, larger than the $X(3872)$ width estimation calculated in the work
  of $\Gamma(X(3872)) < 1.2$ MeV at 90\% CL.

 \item  Ablikim {\it et al}~\cite{Ablikim:2013dyn} analyze the angular distribution of the radiative photon in the $e^+e^-$ CM
 frame and the $\pi^+\pi^-$ invariante mass distribution. For the $X(3872)$ signal, they select events 
 between $3.86  {\rm GeV} < M(\pi^+ \pi^- J/\psi) < 3.88$ GeV.
 The mass resolution of the detector is estimated from fits to the $\psi(3686)$ signal, obtaining
 $\sigma = (1.14 \pm 0.07)$ MeV/c$^2$. Results are shown with an energy bin of $\Delta m=3$ MeV.

\item In Ref.~\cite{Aaij:2011sn}, Aaij {\it et al} study the inclusive production of the $X(3872)$ in pp 
collisions at $\sqrt{s}=7$ TeV. Candidates are selected within a $\pm 3\sigma$ energy window, where the 
mass resolution is estimated as $\sigma = (3.33 \pm 0.08)$ MeV for the $X(2872)$, so $\Delta M \simeq 20$ MeV. 
The energy bin used in the work is $\Delta m=2$ MeV.

\end{itemize}

\end{document}